\documentclass[prb,twocolumn]{revtex4-2}
\usepackage{graphicx,bm,color}

\newcommand{\dy}{\displaystyle}
\newcommand{\bfr}{{\bf r}}
\newcommand{\bfA}{{\bf A}}
\newcommand{\CA}{{\cal A}}
\newcommand{\bfE}{{\bf E}}
\newcommand{\bfj}{{\bf j}}
\newcommand{\bfk}{{\bf k}}
\newcommand{\bfq}{{\bf q}}
\newcommand{\bfG}{{\bf G}}
\newcommand{\bff}{{\bf f}}
\newcommand{\bfF}{{\bf F}}
\newcommand{\bfzero}{{\bf 0}}
\newcommand{\uuH}{\underline{\underline{\hat H}}}
\newcommand{\tchi}{\tilde \chi}

\begin{document}

\title{Stability of the long-range corrected exchange-correlation functional and the Proca procedural functional in time-dependent density-functional theory}
\author{Jared R. Williams}
\affiliation{Department of Physics and Astronomy, University of Missouri, Columbia, Missouri 65211, USA}

\author{Carsten A. Ullrich}
\email{ullrichc@missouri.edu}
\affiliation{Department of Physics and Astronomy, University of Missouri, Columbia, Missouri 65211, USA}

\date{\today }

\begin{abstract}
Excitonic effects in the optical absorption spectra of solids can be described with time-dependent density-functional theory (TDDFT)
in the linear-response regime, using a simple class of approximate, long-range corrected (LRC) exchange-correlation functionals.
It was recently demonstrated that the LRC approximation can also be employed in real-time TDDFT to describe exciton dynamics.
Here, we investigate the numerical stability of the time-dependent LRC approach using a two-dimensional model solid. It is found that
the time-dependent Kohn-Sham equation with an LRC vector potential becomes more and more prone to instabilities for increasing exciton binding energies. The origin of these instabilities is traced back to time-averaged violations of the zero-force theorem, which leads to
a simple and robust numerical stabilization scheme. This explains and justifies a recently proposed method by Dewhurst {\em et al.} [Phys. Rev. B {\bf 111}, L060302 (2025)]
to stabilize the LRC vector potential, known as the Proca procedural functional.
\end{abstract}

\maketitle

\section{Introduction}
The optical properties of insulators and semiconductors -- or, more precisely, the electronic excitations close to the band gap -- are strongly influenced by
excitonic effects \cite{Yu2010}. Excitons are often visualized as bound electron-hole pairs which can be described by a simple hydrogen-like Schr\"odinger equation
\cite{Haug}, but in reality the dielectrically screened electron-hole interaction which causes excitonic binding is a complex many-body phenomenon \cite{Bechstedt2015,Reining2016}. The standard theoretical approach to describe excitons is via the Bethe-Salpeter equation (BSE), often in combination
with electronic band structures obtained using the GW method \cite{Onida2002}. An alternative to GW-BSE is time-dependent density-functional theory (TDDFT)
\cite{Runge1984,Ullrich2012}. There are many studies in the literature which show that TDDFT, in the frequency-dependent linear-response regime,
can produce optical spectra in solids that capture excitonic effects with useful accuracy
\cite{Reining2002,Botti2007,Ullrich2015,Turkowski2017,Sun2019,Wing2019,Sun2020,Cavo2020,Tal2020,Gomez2023,Gomez2024}.

A major advantage of TDDFT is that it can be applied not only in the frequency-dependent linear-response regime, but also in the real-time regime, which allows
the description of ultrafast linear or nonlinear phenomena. The GW-BSE approach can also be extended into the real-time regime, but at a significantly higher
computational cost \cite{Attaccalite2011,Jiang2021,Chan2021,Chan2023,Perfetto2022,Reeves2024}. In recent years, many applications of real-time TDDFT for solids have come
forward  \cite{Bertsch2000,Yabana2012,Yamada2019,Tancogne2017,Lian2018,Pemmaraju2020,Shepard2021,Feng2023,Xu2024,Choi2024,Kononov2022}. The majority of these applications use
semilocal exchange-correlation (xc) functions such as the adiabatic local-density approximation (ALDA), mainly for reasons of computational efficiency.
However, semilocal xc functionals do not capture excitonic effects \cite{Onida2002}.

Recently, it was demonstrated that excitons can be described within real-time TDDFT using a special class of approximations, known as long-range corrected (LRC)
functionals \cite{Sun2021}. LRC functionals were introduced in linear-response TDDFT about two decades ago \cite{Reining2002,Botti2004,Botti2005} and have since then been used in many applications and in many varieties \cite{Sharma2011,Trevisanutto2013,Rigamonti2015,Byun2017b,Byun2020}. It is fair to say that the LRC approach
is the simplest way to simulate excitonic physics within TDDFT \cite{Yang2012}; however, as we will discuss in more detail below, it contains empirical parameters
and assumptions, and produces optical spectra that are in general unsatisfactory, especially for strongly bound excitons \cite{Byun2017b,Byun2020}.
Nevertheless, LRC captures the essence of excitons and is therefore of value, in particular if extended into the real-time regime.

A key observation of Ref. \cite{Sun2021} was that the time-dependent LRC (TDLRC) approach worked well for materials with weakly bound excitons, such as bulk silicon,
but failed for systems with strongly bound excitons due to numerical instabilities. However, it was unclear whether these instabilities
could perhaps be cured using better numerical methods, or whether they are unavoidable features of the TDLRC approach. Dewhurst, Gill, Shallcross, and Sharma (DGSS) \cite{Dewhurst2024}
recently proposed a simple solution to stabilize the calculations, by including additional terms in the equation of motion that determines the LRC xc vector potential;
they refer to their approach as Kohn-Sham-Proca equation,  which involves a procedural definition of the xc functional. While this appears to be
a successful cure of the instability problem, the physical justification of this approach is not completely clear.

The purpose of this paper is to carry out a careful and detailed formal and numerical analysis of the TDLRC approach and its numerical
behavior. To avoid costly calculations, we use a simple two-dimensional (2D) model solid which captures all the important physical features
of a real periodic solid but is much simpler and numerically much less expensive to treat. We analyze in detail how the LRC functional produces excitons in real-time TDDFT
and what causes the numerical instabilities. It turns out that the source of the problem is a time-averaged violation of the zero-force theorem of TDDFT, and
the method proposed by DGSS \cite{Dewhurst2024} ultimately derives its justification from this, as we will show.

The paper is organized as follows. Section \ref{Sec:2} contains the necessary theoretical background, covering real-time TDDFT for light-matter interactions
in real and reciprocal space and the definition of the LRC vector potential. In Sec. \ref{Sec:3} we discuss our 2D model solid and give some
numerical details. Results and discussions will be presented in Sec. \ref{Sec:4}, and we give our conclusions in Sec. \ref{Sec:5}.
Additional material is presented in two Appendices.

\section{Theoretical Background}\label{Sec:2}

In the following, we limit the discussion to electronic systems that are nonmagnetic and nonrelativistic (i.e., no spin-orbit coupling),
so that the electronic spin does not need to be treated explicitly. Furthermore, we assume an all-electron description, i.e., we do not need to include pseudopotentials
in our discussion. Atomic units ($\hbar = e = m = 4\pi \varepsilon_0$) are used throughout.

\subsection{Real-time TDDFT for light-matter interaction}

We consider situations where an $N_e$-electron system is initially in the ground state associated with a static external
potential $v(\bfr)$, assumed to be the electrostatic potential of fixed nuclei or a fixed model potential.
At time $t_0$, a spatially uniform vector potential $\bfA(t)$ is switched on. This vector potential is related to a
uniform, time-dependent electric field according to $d\bfA(t)/dt = \bfE(t)$. In this way, the interaction between a solid
and a laser pulse can be simulated within the dipole approximation. Of particular interest is the limiting case of an ultrashort ``kick'' of the
form $\bfE_{\rm kick}(t) = \bfE_0 \delta(t-t_0)$, which yields a vector potential with a step-function time dependence:

\begin{equation}\label{kick}
    \bfA_{\rm kick}(t) = \bfE_0 \theta(t-t_0) \:.
\end{equation}

The system subsequently evolves in time, governed by the time-dependent Kohn-Sham (TDKS) equation:
\begin{eqnarray}\label{TDKS}
    i \frac{\partial}{\partial t} \psi_j(\bfr,t) &=&
    \left[ \frac{1}{2}\left(\frac{\nabla}{i} + \bfA(t) + \bfA_{\rm xc}(t)\right)^2 + v(\bfr)\right. \nonumber\\
    &&  {}+ v_{\rm H}(\bfr,t) + v_{\rm xc}(\bfr,t)\bigg] \psi_j(\bfr,t) \:.
\end{eqnarray}

The Hartree potential $v_{\rm H}(\bfr,t) = \int d\bfr' n(\bfr',t)/|\bfr-\bfr'|$ and the xc potential $v_{\rm xc}[n](\bfr,t)$
are both functionals of the time-dependent density, given by
\begin{equation}
    n(\bfr,t) = \sum_j^{N_e} |\psi_j(\bfr,t)|^2 .
\end{equation}
We assume here that $v_{\rm xc}(\bfr,t)$ has the same functional form as the xc potential in the static Kohn-Sham equation that was used to calculate the initial ground state, i.e., we use the adiabatic approximation.

The xc vector potential $\bfA_{\rm xc}(t)$, on the other hand, is here assumed to be $\bfr$-independent, purely dynamical,
and constructed as a functional of the total current density \cite{Ullrich2012,currentnote},
\begin{equation}
    \bfj(\bfr,t) = 2 \Im \sum_j^{N_e}\psi_j^*(\bfr,t)\nabla \psi_j(\bfr,t) + n(\bfr,t)[ \bfA(t) + \bfA_{\rm xc}(t)].
\end{equation}
The specific form of $\bfA_{\rm xc}(t)$ used in this paper will be discussed below.

\subsection{Reciprocal space formulation}

We now specifically consider periodic solids with time-dependent Kohn-Sham orbitals of the form
\begin{equation}
    \psi_{l\bfk}(\bfr,t) = e^{i\bfk\cdot\bfr} u_{l\bfk}(\bfr,t),
\end{equation}
where the lattice-periodic Bloch functions $u_{l\bfk}$ are expanded in a plane-wave basis:
\begin{equation}
    u_{l\bfk}(\bfr,t) = \sum_\bfG C_{l,\bfk-\bfG}(t) e^{-i\bfG\cdot \bfr} \:.
\end{equation}
Here, $l$ is the band index and $\bfk$ is a wave vector in the first Brillouin zone.
The initial ground state is obtained by solving the static Kohn-Sham equation,
\begin{equation}\label{KS}
    \left(\frac{1}{2}(\bfk-\bfG)^2 - \varepsilon_{l\bfk}\right)C^{(0)}_{l,\bfk-\bfG} + \sum_{\bfG'}U_{\bfG'-\bfG}^{(0)}C^{(0)}_{l,\bfk-\bfG'} = 0.
\end{equation}
Here, $\varepsilon_{l\bfk}$ is the Kohn-Sham band structure, the initial states follow from $C_{l,\bfk-\bfG}^{(0)} = C_{l,\bfk-\bfG}(t_0^-)$, and $U_\bfG^{(0)}$ are the Fourier coefficients of the lattice-periodic total potential $v(\bfr) + v_{\rm H}(\bfr,t_0^-) + v_{\rm xc}(\bfr,t_0^-)$. By $t_0^-$ we denote a time infinitesimally before the time $t_0$ when the perturbation is switched on.

The TDKS equation, Eq. (\ref{TDKS}), then transforms into
\begin{eqnarray}\label{TDKSk}
    i \frac{\partial}{\partial t} C_{l,\bfk-\bfG}(t)
    &=&
    \frac{1}{2}\left[\bfk - \bfG + \bfA(t) + \bfA_{\rm xc}(t)\right]^2 C_{l,\bfk-\bfG}(t) \nonumber\\
    &+&
    \sum_{\bfG'}U_{\bfG'-\bfG}(t)C_{l,\bfk-\bfG'}(t).
\end{eqnarray}
The time-dependent density is given by
\begin{equation}
    n_\bfG(t) = 2\sum_l^{N/2} \sum_\bfk\sum_{\bfG'}C_{l,\bfk-\bfG'}(t) C^*_{l,\bfk-\bfG-\bfG'}(t) \:,
\end{equation}
and the total macroscopic current density is given by
\begin{equation}\label{jmac}
    \bfj_\bfzero(t) = 2\sum_l^{N/2} \sum_\bfk\sum_{\bfG'}\bfG' |C_{l,\bfk-\bfG'}(t)|^2 + N[\bfA(t) + \bfA_{\rm xc}(t)].
\end{equation}
Here, $N$ is the number of electrons per unit cell (an even number), and we assume that our system is an insulator or a semiconductor, with all bands either fully occupied or completely empty. Thus, $N/2$ is the number of occupied bands.

The total time-dependent scalar potential is given by $U_\bfG(t) = v_\bfG + v_\bfG^{\rm H}(t) + v_\bfG^{\rm xc}(t)$. The Hartree potential in reciprocal space is defined as
\begin{equation}\label{vH}
    v_{\bfG}^{\rm H}(t) =w_\bfG n_\bfG(t)\:,\quad \bfG\ne 0 \:,
\end{equation}
where $w_\bfG$ is the Fourier transform of the Coulomb potential, which, for arbitrary wave vectors $\bfq$, is given by
\begin{equation}\label{w}
    w_{\bfq} = \left\{ \begin{array}{cc} \dy \frac{4\pi }{q^2} & \mbox{(3D)}  \\[3mm] \dy \frac{2\pi}{q} & \mbox{(2D).}
    \end{array}\right.
\end{equation}
The diverging $\bfG=0$ component of the Hartree potential is omitted in Eq. (\ref{vH}) since it is canceled by the positive background charge.
The xc potential is given by
\begin{equation}\label{VXCG}
    v_\bfG^{\rm xc}(t) = \frac{1}{\cal V} \int_{\rm cell} d\bfr \:e^{-i\bfG\cdot\bfr}v _{\rm xc}(\bfr,t) \:,
\end{equation}
where $\cal V$ is the unit cell volume and $v _{\rm xc}(\bfr,t)$ is an approximate semilocal xc potential. However, rather than directly evaluating Eq. (\ref{VXCG}) it is much more efficient to start from $v _{\rm xc}(\bfr,t) = \sum_\bfG e^{i\bfG\cdot\bfr}v_\bfG^{\rm xc}(t)$, which in practice is a finite sum over $\bfG$-vectors. Thus, we only need to evaluate $v _{\rm xc}(\bfr,t)$ at a finite number of sampling points $\bfr$ (the same as the number of $\bfG$-vectors), and can then obtain $v_\bfG^{\rm xc}(t)$ using a linear equation solver.

\subsection{LRC vector potential}

In linear-response TDDFT, the LRC xc kernel is defined as \cite{Reining2002,Byun2017b}
\begin{equation}\label{fxcLRC}
    f_{\rm xc}^{\rm LRC}(\bfr,\bfr')= -\frac{\alpha}{4\pi} \frac{1}{|\bfr - \bfr'|} \:,
\end{equation}
where $\alpha>0$ is an adjustable constant. The LRC xc kernel has been demonstrated to be capable of producing excitonic peaks in optical absorption spectra \cite{Byun2020}. We would now like to use it in a real-time TDDFT calculation.

The time-dependent density can always be written as \begin{equation}
    n(\bfr,t) = n_{\rm gs}(\bfr) + \delta n(\bfr,t) ,
\end{equation} where $n_{\rm gs}(\bfr)$ is the ground-state density and $\delta n(\bfr,t)$ is the density response, which is usually assumed to be small compared to $n_{\rm gs}(\bfr)$.
We then define a time-dependent LRC potential that only depends on the density response, \begin{equation}
    v_{\rm xc}^{\rm LRC}(\bfr,t) = -\frac{\alpha}{4\pi} \int d\bfr' \frac{\delta n(\bfr',t)}{|\bfr - \bfr'|} \:.
\end{equation}
Transformed into reciprocal space, we get
\begin{equation}
    v_\bfG^{\rm LRC}(t) = -\frac{\alpha}{4\pi} w_\bfG \delta n_\bfG(t) \:.
\end{equation} This expression becomes ill defined for $\bfG=\bfzero$: because of charge conservation we have $\delta n_{\bfG = \bfzero}(t)=0$, and at the same time $w_{\bfG = \bfzero}$ diverges. This is a problem since the $\bfG=\bfzero$ component of $f_{\rm xc}^{\rm LRC}$ -- the head of the xc kernel matrix -- is known to be the dominant contributor for excitonic effects in linear-response TDDFT.

The solution to this problem is to treat the $\bfG=\bfzero$ part as a vector potential \cite{Maitra2003}. Using the gauge relation between scalar and vector potentials, \begin{equation}\label{18}
    \frac{\partial}{\partial t} \bfA(\bfr,t) = -\nabla v(\bfr,t) \:,
\end{equation}
and the continuity equation,
\begin{equation}\label{19}
    \nabla \cdot \bfj(\bfr,t) = -\frac{\partial}{\partial t}n(\bfr,t) \:,
\end{equation}
we obtain the following expression for the LRC vector potential:
\begin{equation}
    \bfA_{\rm xc}^{\rm LRC}(\bfr,t) = -\frac{\alpha}{4\pi} \int_{t_0}^t dt' \int_{t_0}^{t'} dt'' \nabla \int d\bfr' \frac{\nabla' \cdot \bfj(\bfr',t'')}
    {|\bfr - \bfr'|} \:.
\end{equation}
In reciprocal space, this gives, for general $\bfq$,
\begin{equation}
    \bfA^{\rm LRC}_{\rm xc, \bfq}(t) = \alpha\int_{t_0}^t dt' \int_{t_0}^{t'} dt'' w_\bfq \bfq [ \bfq\cdot \bfj_\bfq(t'')].
\end{equation} We set $\bfq = \bfk + \bfG$ and consider the dominant contribution at $\bfG=\bfzero$, letting $\bfk\to \bfzero$ along the direction of the macroscopic current density
$\bfj_{\bfzero}(t)$. In 3D, this gives
\begin{equation}\label{A3D}
    \bfA^{\rm LRC,3D}_{\rm xc,\bfzero}(t) = \alpha\int_{t_0}^t dt' \int_{t_0}^{t'} dt'' \bfj_{\bfzero}(t'') \:.
\end{equation}
In 2D, we obtain
\begin{equation}\label{A2D}
    \bfA^{\rm LRC,2D}_{\rm xc,\bfzero}(t) = \lim_{k\to 0}\frac{\alpha k}{2}\int_{t_0}^t dt' \int_{t_0}^{t'} dt'' \bfj_{\bfzero}(t'') \:.
\end{equation}
This means that in 2D we will consider the optical response at a small but finite wavevector $\bfk$.
For numerical evaluation it is preferable to rewrite Eqs. (\ref{A3D}) and (\ref{A2D}) in the form of a second-order differential equation:
\begin{eqnarray}\label{LRC-vec-pot-diffeq-3D}
    \frac{d^2}{dt^2}\bfA^{\rm LRC,3D}_{\rm xc,\bfzero}(t)  &=& \alpha  \bfj_{\bfzero}(t)\\
    \frac{d^2}{dt^2}\bfA^{\rm LRC,2D}_{\rm xc,\bfzero}(t)  &=&  \lim_{k\to 0}\frac{\alpha k}{2} \bfj_{\bfzero}(t) \:.
    \label{LRC-vec-pot-diffeq-2D}
\end{eqnarray}

\subsection{Modified LRC vector potential: the DGSS-Proca approach}

The equations of motion (\ref{LRC-vec-pot-diffeq-3D}) and (\ref{LRC-vec-pot-diffeq-2D}) for the LRC vector potential are nonlinear:
the macroscopic current density
$\bfj_{\bfzero}$, Eq. (\ref{jmac}), depends on the xc vector potential both implicitly, via $C_{l,\bfk-\bfG}(t)$, and explicitly, via the diamagnetic
contribution $N\bfA_{\rm xc}(t)$. For finite $\alpha$ this means that the solution experiences reinforcement through feedback, which is exactly what
creates the excitonic resonances that are absent in RPA or ALDA, but it can also lead to catastrophic failure in the form of numerical instabilities. How this explicitly works will be discussed in detail below, see in particular Sec. \ref{Sec:4C}.

Before proceeding, a brief clarification. The ALDA is also nonlinear: in fact, exciton-like resonances can be generated using
only the ALDA xc potential in Eq. (\ref{TDKS}), but artificially scaled up by a large prefactor. This exciton arises entirely
through local-field  effects, since ALDA does not have any long-range electron-hole interactions.
This scaling approach was used in Ref. \cite{Sun2021} as an alternative to TDLRC, and shown to be numerically stable;
however, it is physically not well motivated. On the other hand, the key feature of LRC is that it is both long-ranged and nonlinear:
this is why LRC gives excitons for the right physical reason \cite{Reining2002}, at the cost of possible numerical instabilities when
going beyond linear response.

To make the LRC vector potential better behaved, DGSS  proposed to incorporate additional
terms into the equation of motion for the LRC vector potential, as follows \cite{Dewhurst2024}:
\begin{equation}\label{mLRC}
    \frac{d^2}{dt^2}\bfA^{\rm DGSS}_{\rm xc,\bfzero}(t) + \beta\frac{d}{dt}\bfA^{\rm DGSS}_{\rm xc,\bfzero}(t)
    + \gamma \bfA^{\rm DGSS}_{\rm xc,\bfzero}(t) = \alpha  \bfj_{\bfzero}(t) \:,
\end{equation}
where $\beta$ and $\gamma$ are adjustable real parameters.  We note that DGSS have named the combination of Eq. (\ref{mLRC}) and the TDKS equation (\ref{TDKS}) the Kohn-Sham-Proca scheme.

$\bfA^{\rm LRC}_{\rm xc,\bfzero}(t)$ and $\bfA^{\rm DGSS}_{\rm xc,\bfzero}(t)$ are procedural xc functionals since they are defined
via equations of motion which have to be time propagated together with the TDKS equation \cite{Dewhurst2024}.
Clearly, Eq. (\ref{mLRC}) has the same form as a damped driven harmonic oscillator, where $\gamma$ is the spring
constant divided by the mass and $\beta$ is the damping parameter; we will assume here that $\beta < 2\gamma$ (underdamped case).
Setting $\Omega_0 = \sqrt{\gamma}$ we can then identify the characteristic oscillator frequency $\Omega = \sqrt{\Omega_0^2 - \beta^2/4}$.

The DGSS-Proca approach is a very effective means to simulate strongly bound excitons. It introduces an additional (spurious) resonance at $\Omega$, which in practice can be chosen at frequencies much smaller than the band gap and hence sufficiently far away from the excitonic resonance, as we shall demonstrate below.
The stabilizing effect of the additional terms in the equation of motion (\ref{mLRC}), which prevents runaway numerical behavior,
will be analyzed in detail in the next Sections.

It is straightforward to show from Eq. (\ref{mLRC}), and using Eqs. (\ref{18}) and (\ref{19}),
 that in the linear-response regime the DGSS scheme leads to a frequency-dependent xc kernel:
\begin{equation}\label{fxcLRCprime}
    f_{\rm xc}^{\rm DGSS}(\bfr,\bfr',\omega)= \frac{\omega^2 f_{\rm xc}^{\rm LRC}(\bfr,\bfr')}{\omega^2 +i\omega \beta -\gamma}\:.
\end{equation}
It is interesting to compare this to the dynamical xc kernel proposed by
Botti {\em et al.} \cite{Botti2005}, which has the form
$f_{\rm xc}^{\rm dyn}(\bfr,\bfr',\omega) =  -(\alpha_0 + \alpha_2 \omega^2)/(4\pi|\bfr - \bfr'|)$.
This corresponds to the following equation of motion:
\begin{equation}\label{dyn}
    \frac{d^2}{dt^2}\bfA^{\rm dyn}_{\rm xc,\bfzero}(t)  = \alpha_0  \bfj_{\bfzero}(t) - \alpha_2 \frac{d^2}{dt^2} \bfj_{\bfzero}(t) \:.
\end{equation}
In the linear response regime, the additional parameter $\alpha_2$ in $f_{\rm xc}^{\rm dyn}$ can lead to optical absorption spectra in better agreement with
experiment \cite{Botti2005} than pure LRC. However, we have tested the equation of motion Eq. (\ref{dyn}) and found that it still suffers from instabilities
due to unmitigated feedback. In Ref. \cite{Reshetnyak2019} it was shown that the  dependence of the xc kernel is, in general, more complicated than the
quadratic dependence assumed in $f_{\rm xc}^{\rm dyn}$.

\begin{figure}
  \includegraphics[width=\linewidth]{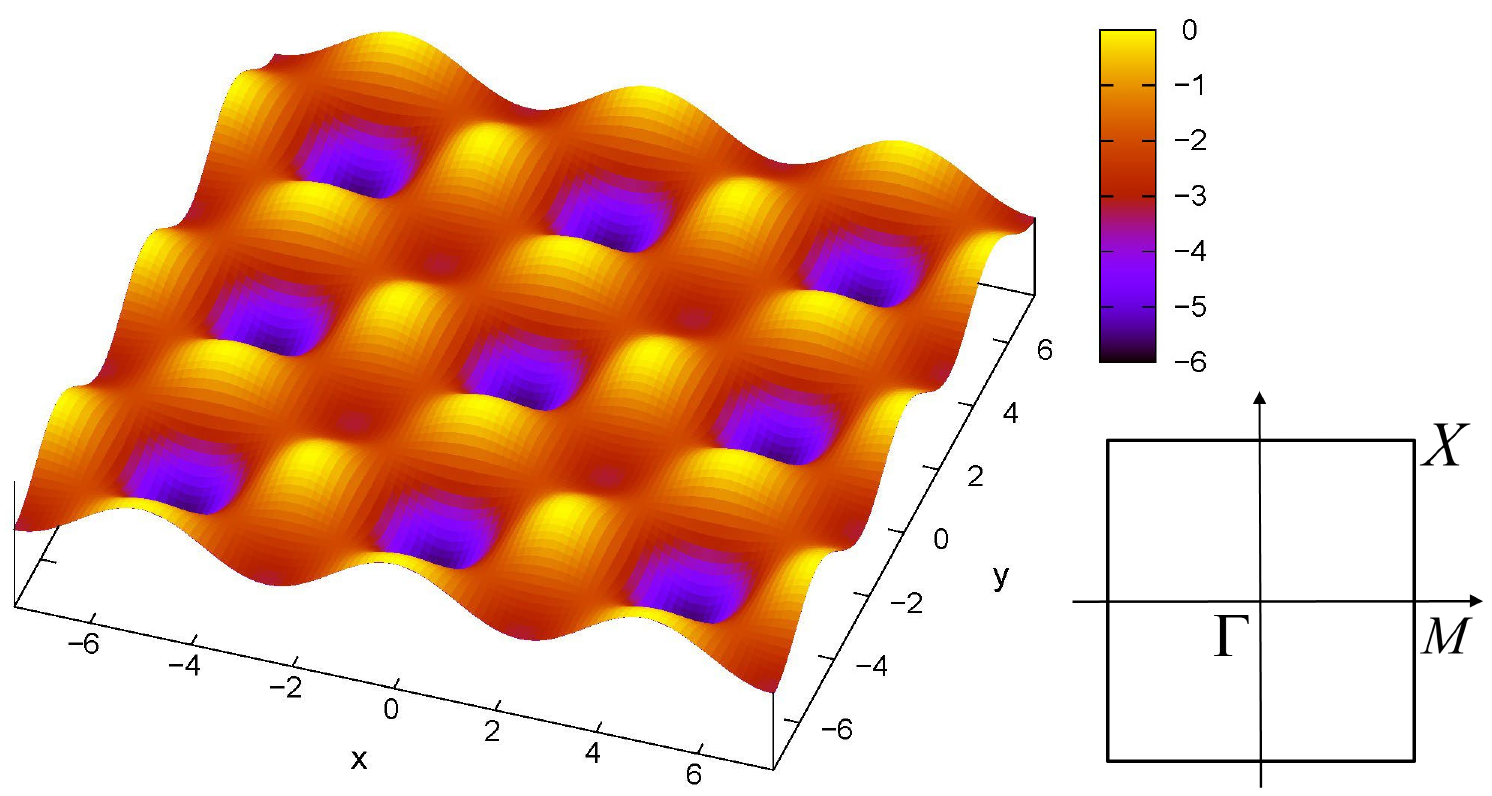}
  \caption{\label{fig1} Left: 2D model potential defined in Eq. (\ref{2Dv}), for $A=1.4$ and $B=0.8$ and lattice constant $c=5$.
  Right: first Brillouin zone of the square lattice.}
\end{figure}

\begin{figure}
  \includegraphics[width=\linewidth]{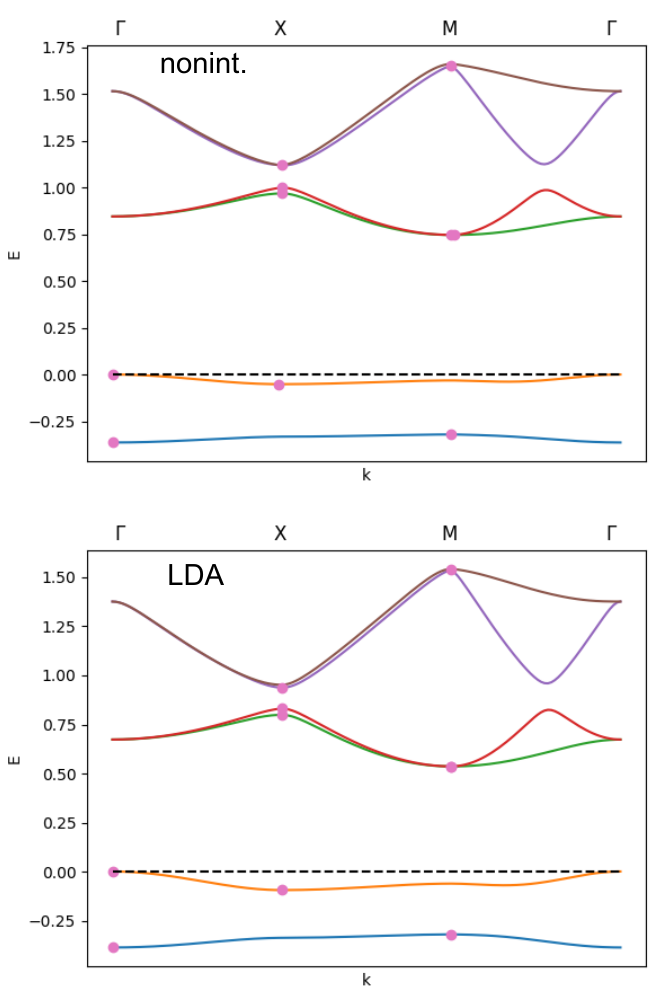}
  \caption{\label{fig2} Electronic band structure of the square lattice, for $A=1.0$, $B=0.9$ and $c=5$, with $N=4$ electrons per unit cell. Top and bottom: noninteracting
  and LDA band structures, respectively. The horizontal dashed lines indicate the Fermi level. The pink dots illustrate band minima and maxima.}
\end{figure}

\section{Model and numerical details}\label{Sec:3}

\subsection{2D model solid}

We consider a 2D periodic solid with an external potential $v(\bfr)$ of the form
\begin{eqnarray}\label{2Dv}
    v(\bfr) &=& -A \left[ \left(\cos\left(\frac{2\pi x}{c}\right)+1\right)\left(\cos\left(\frac{2\pi y}{c}\right)+1\right)\right] \nonumber\\
    &-& B \left[ \left(\cos\left(\frac{2\pi x}{c}\right)- 1\right)\left(\cos\left(\frac{2\pi y}{c}\right)-1\right)\right]
\end{eqnarray}
where $A$ and $B$ are constants. This defines a 2D square lattice with lattice constant $c$ and a ``diatomic'' basis, as illustrated in Fig. \ref{fig1}. The associated reciprocal lattice vectors are given by
\begin{equation}
    \bfG_{n_x, n_y} = \frac{2\pi}{c}\left( \begin{array}{c} n_x \\n_y \end{array}\right), \qquad
    n_x,n_y = 0,\pm 1, \pm2, \ldots
\end{equation}
The Fourier transform of $v(\bfr)$ from Eq. (\ref{2Dv}) is
\begin{eqnarray}
    v_{\bfG_{n_x,n_y}} &=& -\frac{(A-B)}{2}\: [\delta_{n_x,\pm 1}\delta_{n_y,0}  + \delta_{n_x,0}\delta_{n_y,\pm 1} ] \nonumber\\
    &&
    -\frac{(A+B)}{4}\: \delta_{n_x,\pm 1}\delta_{n_y,\pm 1} \:.
\end{eqnarray}
The first Brillouin zone associated with this lattice is a square, as illustrated in Fig. \ref{fig1}. By adjusting the potential parameters $A$ and $B$ and the number of electrons per unit cell, $N$, we can make this system metallic, insulating or semiconducting, with direct or indirect gap. In the following we choose
$N=4$ throughout.

The smooth cosine shape of the potential wells means that a simple plane-wave expansion of the Kohn-Sham Bloch functions will be suitable. In the following, we let $n_x,n_y = 0,\pm 1, \ldots, \pm n_G$, i.e., we include $(2n_G+1)^2$ reciprocal lattice vectors $\bfG_{n_x, n_y}$. We choose a uniform $\bfk$-point grid with $(n_k)^2$ points in each quadrant of the square Brillouin zone (the advantage of this choice is ease of implementation). If only time-dependent scalar potentials are present, then the symmetry of
the problem allows us to consider only $\bfk$-points in one quadrant of the Brillouin zone. However, if time-dependent vector potentials are present then
$\bfk$-points in all four quadrants of the Brillouin zone are needed, which of course increases the numerical effort.

The numerical results presented in the following have been obtained with a homemade Python code, which is publicly available \cite{Code}.
The code has the capabilities of calculating the band structure of the 2D model solid, the frequency-dependent dielectric function, and solving the TDKS equation
including LRC.

Figure \ref{fig2} shows the electronic band structure of the square lattice for $A=1.0$, $B=0.9$, and $c=5$, comparing noninteracting calculation ($v_{\rm H}=v_{\rm xc}=0$) and LDA.
As can be seen from the band structures, for these parameters the system is a strong insulator with an indirect band gap of 0.75 a.u. (noninteracting) and 0.54 a.u. (LDA).

\subsection{Time propagation algorithms}

\subsubsection{Time-dependent Kohn-Sham equation}

For the numerical propagation of the TDKS equation, Eq. (\ref{TDKSk}), we discretize the time coordinate $t$ with a uniform time step $\Delta t$. The TDKS equation decouples into $n_Q(n_k)^2$ independent equations for all $\bfk$-points considered ($n_Q$ is the number of quadrants of the Brillouin zone included), times the number of occupied bands, $l=1,\ldots,N/2$. For each choice of $l,\bfk$ we arrange the coefficients $C_{l,\bfk-\bfG}(t)$ into a vector $\vec C_{l,\bfk}(t)$ of length $(2n_G+1)^2$, which satisfies the matrix equation
\begin{equation}
    i\frac{\partial}{\partial t}\vec C_{l,\bfk}(t) = \uuH_\bfk(t) \vec C_{l,\bfk}(t)  \:,
\end{equation}
where the Hamiltonian matrix is defined in Eq. (\ref{TDKSk}).

We implemented two algorithms \cite{Castro2004} to achieve the propagation step $t_j \to t_{j+1}$. The first is the implicit Crank-Nicolson algorithm, where
\begin{equation}\label{CN}
    \vec C_{l,\bfk}(t_{j+1}) =
    \frac{\underline{\underline{1}} - \frac{i \Delta t}{2}\underline{\underline{\hat H}}_\bfk(t_{j+\frac{1}{2}}) }
    {\underline{\underline{1}} + \frac{i \Delta t}{2}\underline{\underline{\hat H}}_\bfk(t_{j+\frac{1}{2}})}\, \vec C_{l,\bfk}(t_{j}) \:.
\end{equation}
Notice that the Hamiltonian must be evaluated at the mid point of the time step, $t_{j+\frac{1}{2}}$. For those parts of the Hamiltonian that depend self-consistently on the density, i.e., the scalar Hartree and xc potentials, this can be achieved using a predictor-corrector scheme (the DGSS vector potential will be treated differently, see below). In practice, we found that no more than two corrector steps are needed; often, a single corrector step is sufficient.

The second algorithm is the exponential mid-point rule:
\begin{equation}\label{EMR}
    \vec C_{l,\bfk}(t_{j+1}) =
    \exp\left\{ - i \Delta t \uuH_\bfk(t_{j+\frac{1}{2}})\right\} \vec C_{l,\bfk}(t_{j}) \:.
\end{equation}
This requires calculating the exponential of a matrix, which is usually avoided due to the computational cost, or done approximately using a Taylor expansion.  Here, however, the matrices $ \uuH_\bfk$ are quite small, and we can afford to implement the exponential mid-point rule numerically exactly using the {\tt Python} function {\tt scipy.linalg.expm} within a predictor-corrector scheme. Both algorithms, Eqs. (\ref{CN}) and (\ref{EMR}), are unitary and time reversible; in most cases we found the exponential mid-point rule to be preferable over Crank-Nicolson.

\begin{figure}
  \includegraphics[width=\linewidth]{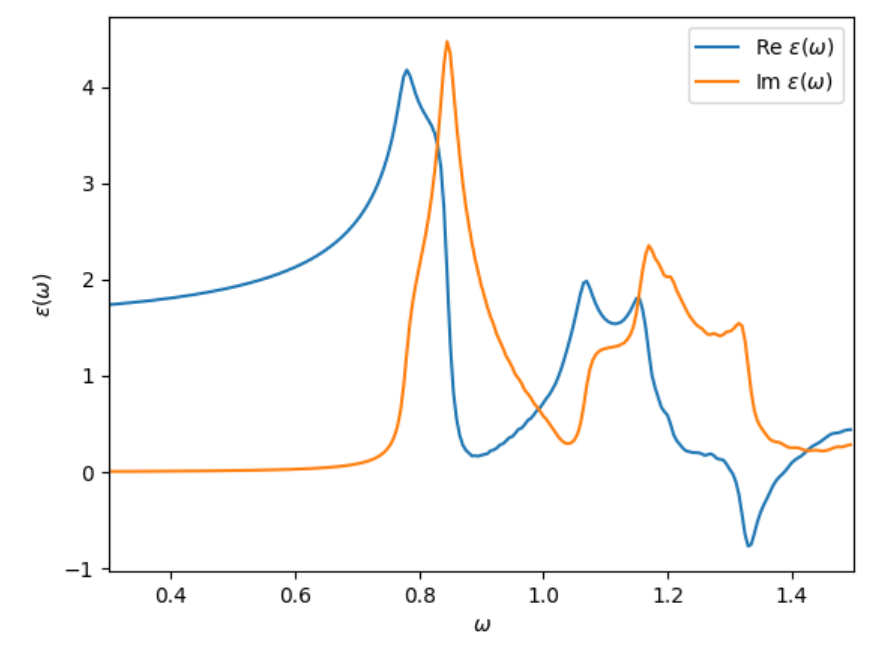}
  \caption{\label{fig3} Real and imaginary part of the macroscopic dielectric function of the noninteracting 2D model solid whose band structure
  is shown in the top panel of Fig. \ref{fig2}. The calculation was done using linear-response TDDFT.}
\end{figure}

\begin{figure}
  \includegraphics[width=\linewidth]{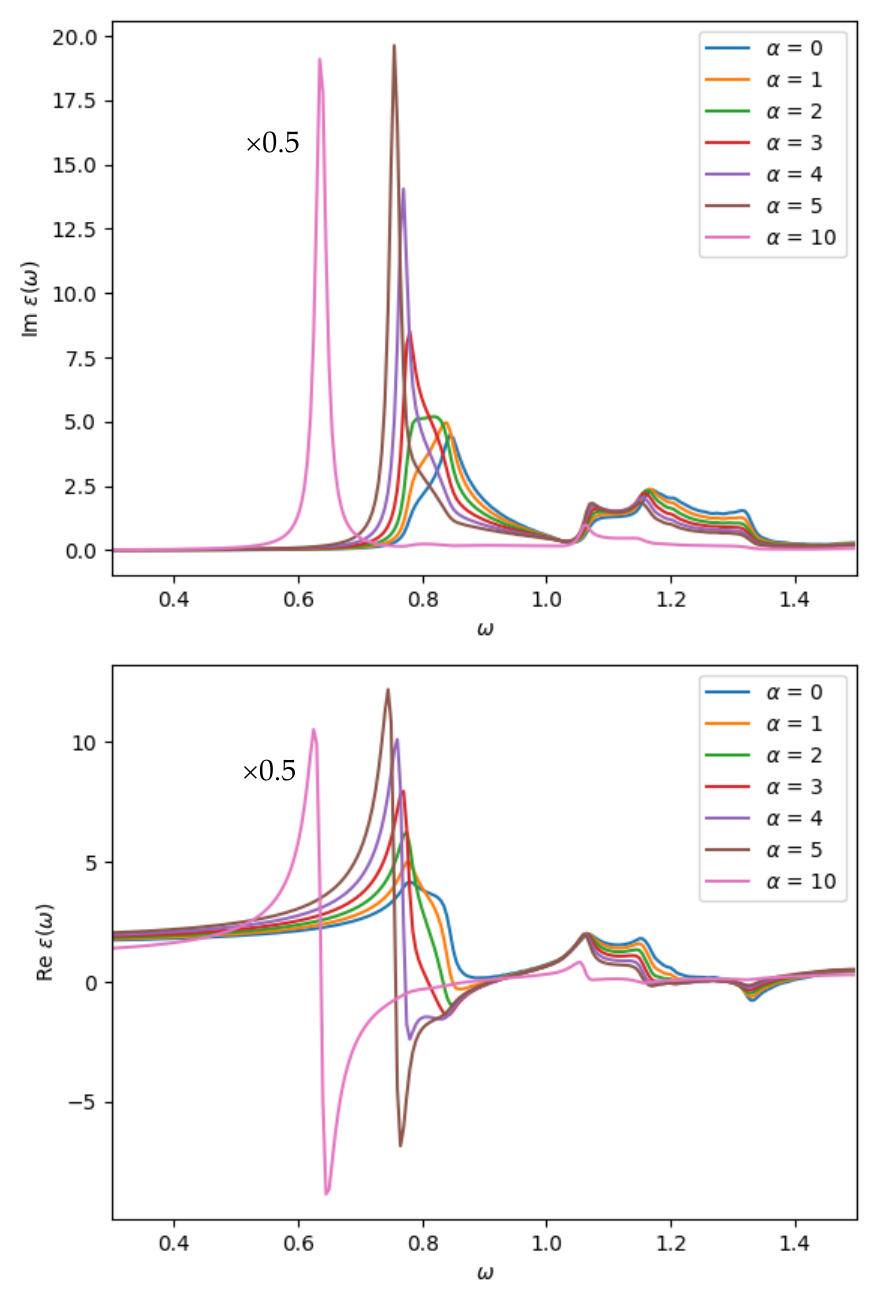}
  \caption{\label{fig4} Real part (top) and imaginary part (bottom) of $\epsilon_{\rm mac}(\omega)$ for the same system as in Fig. \ref{fig3},
  calculated via linear-response TDDFT with the LRC kernel with various values of $\alpha$, as indicated. The increasing strength of the excitonic peak is clearly visible. }
\end{figure}

\subsubsection{DGSS vector potential}

To solve Eq. (\ref{mLRC}) we have implemented a modified version of the St{\o}rmer-Verlet integration method \cite{Press}. In the following, we abbreviate $\bfA^{\rm DGSS}_{\rm xc,\bfzero}(t) \equiv \CA(t)$.

To perform the $t_j \to t_{j+1}$ iteration step we need the vector potential at $\CA(t_{j+\frac{1}{2}})$. We use the
finite-difference forms of the first and second time derivatives,
\begin{eqnarray}
\dot \CA(t_j) &=& \frac{\CA(t_{j+\frac{1}{2}}) - \CA(t_{j-\frac{1}{2}})}{\Delta t}\\
\ddot \CA(t_j) &=& \frac{\CA(t_{j+\frac{1}{2}}) - 2\CA(t_j) + \CA(t_{j-\frac{1}{2}})}{(\Delta t)^2/4}
\end{eqnarray}
and substitute into Eq. (\ref{mLRC}). Solving for $\CA(t_{j+\frac{1}{2}})$ we obtain
\begin{eqnarray}\label{Verlet1}
\CA(t_{j+\frac{1}{2}})
&=&
\Big[ \alpha (\Delta t)^2 \bfj_\bfzero(t_j) + (8-\gamma(\Delta t)^2)\CA(t_j) \nonumber\\
&&
{}+(\beta \Delta t - 4)\CA(t_{j-\frac{1}{2}})\Big](\beta \Delta t + 4)^{-1}.
\end{eqnarray}
$\CA(t_{j-\frac{1}{2}})$ is assumed to be known from the previous iteration step $t_{j-1} \to t_{j}$, but we still need $\CA(t_j)$ [which also enters $\bfj_\bfzero(t_j)$,
see Eq. (\ref{jmac})]. For this we use again finite-difference forms, but now of the form
\begin{eqnarray}
\dot \CA(t_{j-1}) &=& \frac{\CA(t_{j}) - \CA(t_{j-2})}{2\Delta t}\\
\ddot \CA(t_{j-1}) &=& \frac{\CA(t_{j}) - 2\CA(t_{j-1}) + \CA(t_{j-2})}{(\Delta t)^2} \:.
\end{eqnarray}
Substituting into Eq. (\ref{mLRC}) and solving for $\CA(t_j)$ gives
\begin{eqnarray}\label{Verlet2}
\CA(t_{j})
&=&
\Big[ 2\alpha (\Delta t)^2 \bfj_\bfzero(t_{j-1}) + (4-2\gamma(\Delta t)^2)\CA(t_{j-1}) \nonumber\\
&&
{}+(\beta \Delta t - 2)\CA(t_{j-2})\Big](\beta \Delta t + 2)^{-1},
\end{eqnarray}
where $\CA(t_{j-1})$, $\bfj_\bfzero(t_{j-1})$ and $\CA(t_{j-2})$ are assumed to be known from previous time steps.
With $\CA(t_{j})$ obtained in this way we can then get $\CA(t_{j+\frac{1}{2}})$ from Eq. (\ref{Verlet1}).
The initial time propagation step $t_0 \to t_1$ is unproblematic since we can simply set $\CA = 0$ for all $t_j$ with $j\le 0$.

\begin{figure}
  \includegraphics[width=\linewidth]{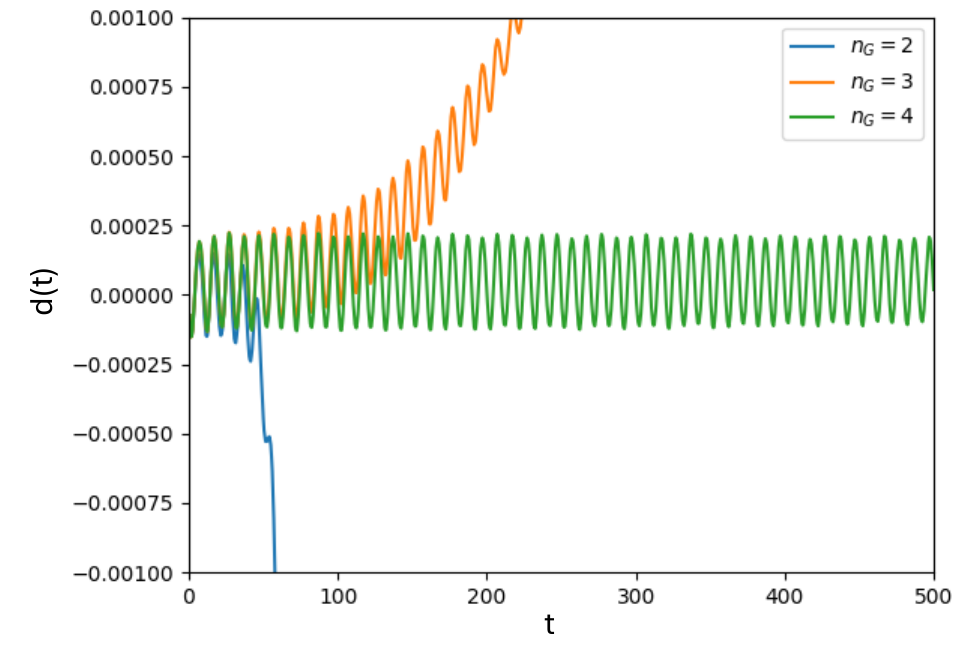}
  \caption{\label{fig5} Time-dependent dipole moment $d(t)$, for TDLRC with $\alpha=5.0$, calculated for three different numbers of $G$-vectors. }
\end{figure}

\begin{figure}
  \includegraphics[width=\linewidth]{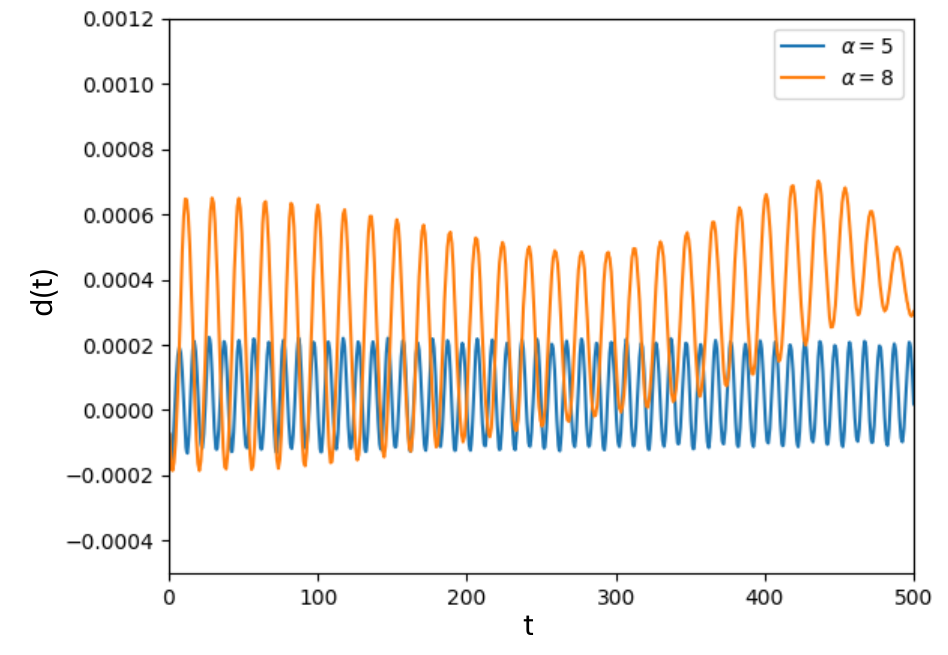}
  \caption{\label{fig6} Time-dependent dipole moment $d(t)$, for TDLRC with $\alpha=5.0$ and $8.0$, calculated with $n_G=4$. }
\end{figure}

\begin{figure*}
  \includegraphics[width=\linewidth]{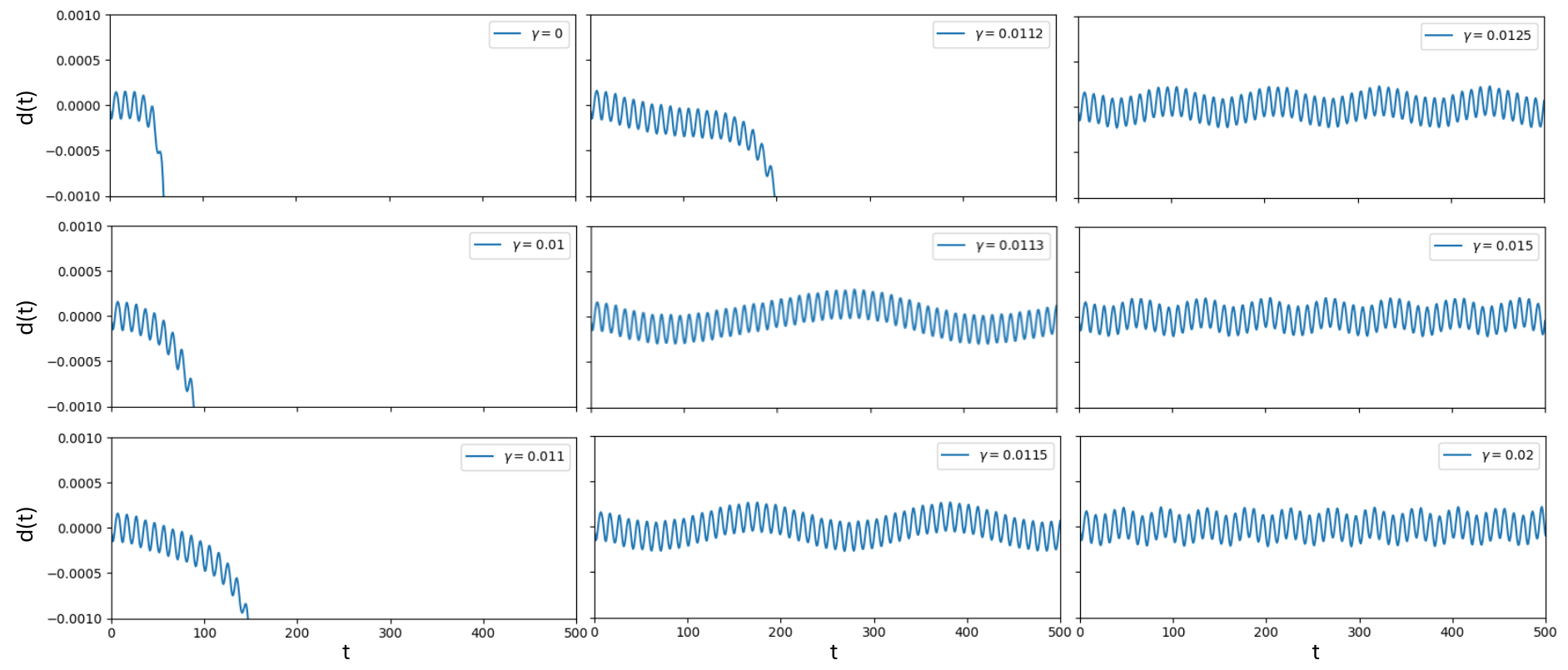}
  \caption{\label{fig7} Time-dependent dipole moment $d(t)$, for DGSS with $\alpha=5.0$, calculated with $n_G=2$. As the parameter $\gamma$ is increased,
  the dipole oscillations stabilize. The threshold is at $\gamma_{\rm thr} = 0.0112$.}
\end{figure*}

\section{Results and discussion}\label{Sec:4}

In the following, we will consider square lattices with $n_k \times n_k$ $\bfk$-points per Brillouin zone quadrant, with $n_k=10$ or 20, and $n_G=2$ or 3, as indicated. All TDDFT results will be for the smallest nonvanishing $\bfk$-vector, i.e., $\bfk = (0.03,0)$ for $n_k=10$
and $\bfk = (0.06,0)$ for $n_k=20$.

Figure \ref{fig2} shows that the noninteracting and LDA band structures are practically identical, apart from the smaller LDA band gap.
We therefore set $v_{\rm H}=v_{\rm xc}=0$ throughout, and include electronic interactions only via the LRC kernel in linear response
or the LRC vector potential in real-time calculations. This does not affect any of our conclusions regarding the performance of LRC.

\subsection{LRC in the linear response regime}\label{Sec:4A}

We begin by calculating the macroscopic dielectric function $\epsilon_{\rm mac}(\omega)$ as discussed in Appendix \ref{App:A}, see Eq. (\ref{emac}).
We choose $n_k=20$ and $n_G=3$ in this Section.
Figure \ref{fig3} shows the real and imaginary parts of $\epsilon(\omega)$ for the noninteracting system and $\alpha=0$.
The onset of absorption is at $\omega = 0.775$ a.u.;
the strong absorption peak slightly above 0.8 a.u. is due to transitions between the highest valence band and the lowest conduction band.
The broader feature around 1.25 a.u. is due to transitions to the next conduction band.

Figure \ref{fig4} shows how the absorption spectrum is modified due to the LRC kernel. With increasing $\alpha$, a very pronounced excitonic peak
starts to develop at the absorption edge. For large enough $\alpha$, the exciton peak moves to energies below the absorption edge: for $\alpha=5$ it is located at 0.755 a.u.,
and at $\alpha=10$ it is at 0.637 a.u., indicating a very strongly bound exciton.
The height of the exciton peak increases very strongly with $\alpha$, which is a known unphysical feature of the LRC kernel:
it is not possible to find an $\alpha$ that produces the correct exciton binding energy and also the correct oscillator strength for a given material
\cite{Byun2017b,Byun2020}.

\subsection{LRC and DGSS in the real-time regime}\label{Sec:4B}

The instability of the TDLRC approach had been first encountered for strongly bound excitons in bulk LiF and in hydrogen chains \cite{Sun2021}.
Let us now illustrate this behavior for our 2D model solid.
We consider the same 2D system as in Fig. \ref{fig2}, using the noninteracting ground state as the initial state. We apply a ``kick'' vector potential of the
form of Eq. (\ref{kick}), with $E_0=0.001$ and directed 45$^{\rm o}$ in the $x-y$ plane. The time propagation is then carried out numerically as described
above, over a time span of $500$ a.u. (which corresponds to 12 fs).

\subsubsection{Diagnosing the instability}

Figure \ref{fig5} shows the time-dependent dipole moment $d(t)$ (see Appendix B) following the weak electric field kick, obtained with TDLRC and $\alpha=5.0$, i.e.,  the
excitonic interactions are moderately strong. Here, the calculations use $n_k=10$ and three different
values of the $G$-vector cutoff, $n_G=2,3,4$, and the time propagation is carried out with a time step $dt=0.5$.
For the $n_G=2$, the dipole moment diverges after just a few oscillations, around $t=75$. The situation improves somewhat at $n_G=3$, but
around $t=200$ the calculation becomes unstable again. At $n_G=4$, the calculation is stable and the system is steadily oscillating.

This behavior is very typical for TDLRC: even for small values of $\alpha$, the time-dependent charge-density oscillations tend to become rapidly unstable;
to some extent, this is a numerical effect which can be mitigated by increasing the number of reciprocal lattice vectors. However, for larger
values of $\alpha$ this becomes increasingly difficult and may require impractical numerical effort.
This is shown in Fig. \ref{fig6}: choosing $n_G=4$, the dipole oscillations are stable for $\alpha=5.0$ but are clearly not stable at $\alpha=8.0$.

\subsubsection{Introducing $\gamma$}

The problem can be fixed by the inclusion of $\gamma$, even if $n_G$ is chosen to be small.
This is demonstrated in Fig. \ref{fig7}, which shows the time-dependent dipole oscillations for $n_G=2$. In TDLRC, the calculation blows up quite rapidly,
around $t=75$. But if we now carry out DGSS calculations with a finite $\gamma$, it can be clearly seen how the dipole oscillations begin to stabilize
as $\gamma$ is increased. There appears to be a rather sharp threshold at $\gamma_{\rm thr}=0.0112$  after which $d(t)$ suddenly becomes stable.
Above this threshold, we observe a slow modulation of the dipole oscillation, with a modulation frequency that increases with $\gamma$. We will discuss this in more detail below.

We point out that the $\beta$-parameter does not appear to play a significant role in stabilizing the dipole oscillations. We
have set $\beta=0$ in all DGSS calculations.

\begin{figure}
  \includegraphics[width=\linewidth]{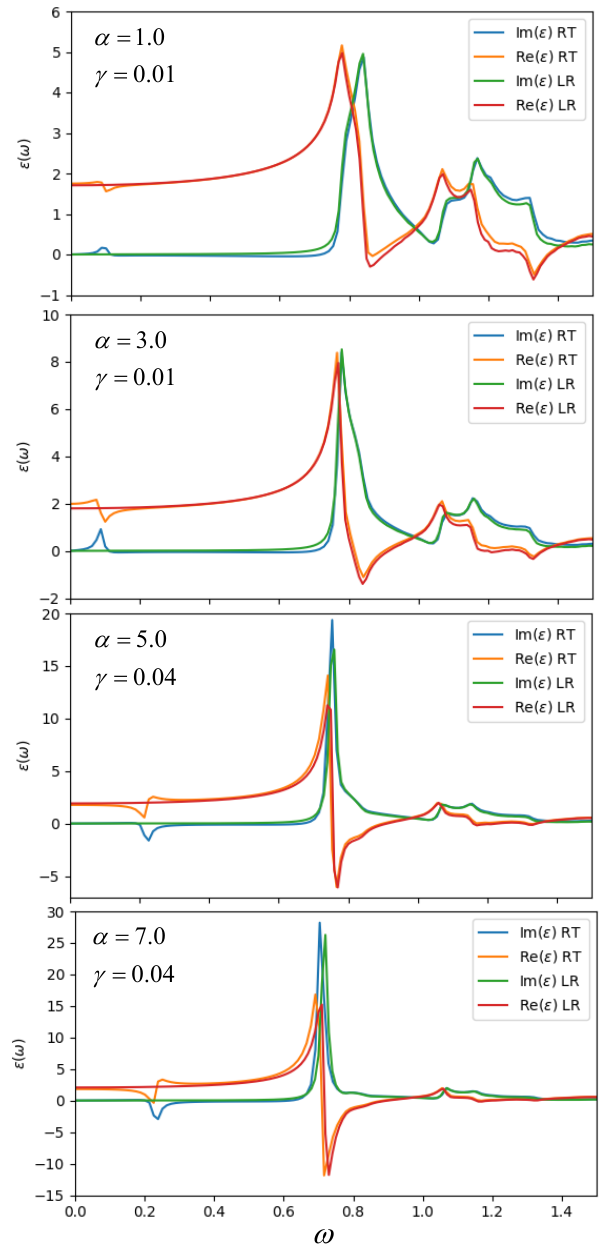}
  \caption{\label{fig8} Comparison of $\epsilon(\omega)$ calculated using DGSS in linear response and with real-time propagation, for different values of $\alpha$. }
\end{figure}

\subsubsection{Comparison between linear response and real time}

To test whether the DGSS approach is compatible with the spectra we obtained in the linear-response regime with LRC (see Section \ref{Sec:4A}) and with DGSS, we consider
a $\bfk$-point grid with $n_k=20$, we choose $n_G=2$, and we carry out the time propagation with the smaller time step $dt = 0.1$ to gain numerical accuracy.
From the time-dependent dipole oscillations we then calculate the optical spectrum following Appendix B. The results are shown in Fig. \ref{fig8}, for
four different values of $\alpha$. To stabilize the DGSS calculations we choose $\gamma=0.01$ for $\alpha = 1$ and 3 and
$\gamma=0.04$ for $\alpha = 5$ and 7. Clearly, there is excellent agreement between the real and imaginary parts of $\epsilon(\omega)$, calculated
using linear response (LR) and real-time (RT).

Closer observation reveals an interesting detail in the real-time spectra: for low frequencies, there are small, spurious features in the real-time spectra that do
not exist in linear response. These features are related to the slow modulations in the $d(t)$ signals that were seen in Fig. \ref{fig7}.

\begin{figure*}
  \includegraphics[width=\linewidth]{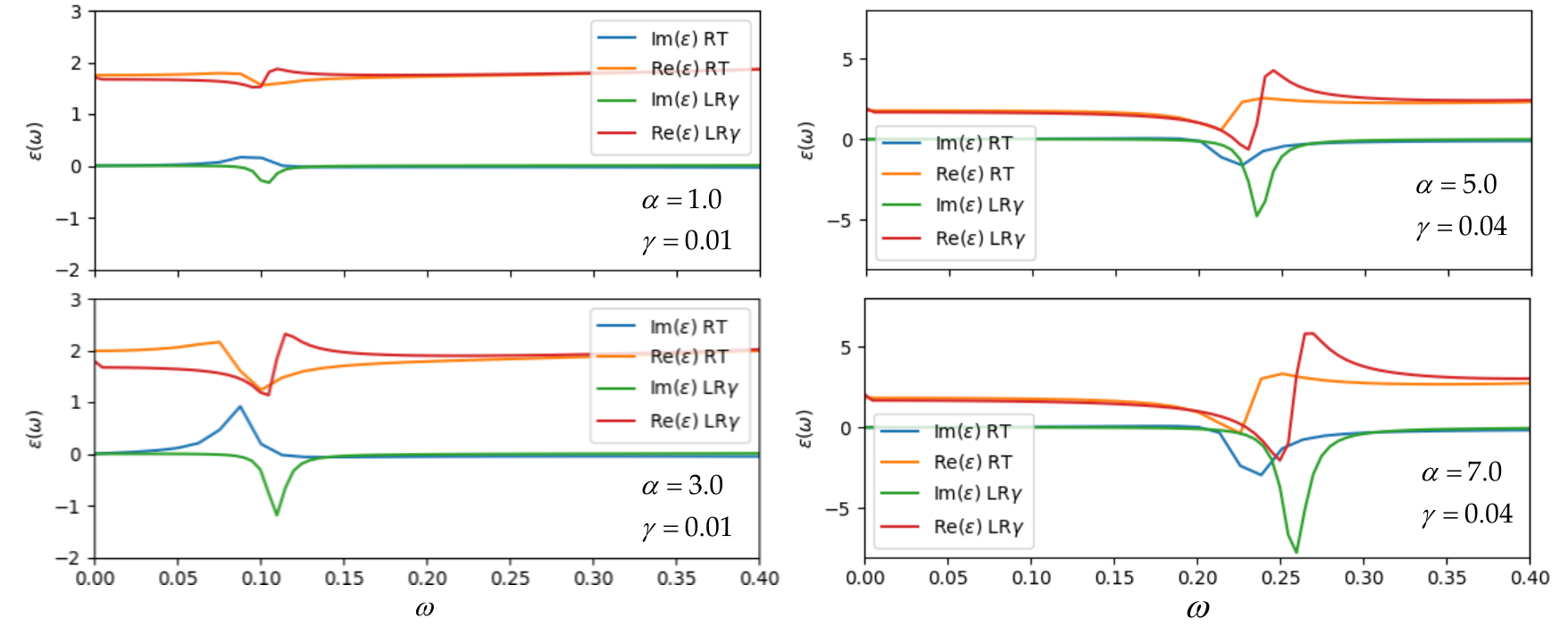}
  \caption{\label{fig9} Closeup of $\epsilon(\omega)$ in the low-frequency region, comparing real-time DGSS with linear-response DGSS using
  the xc kernel of Eq. (\ref{fxcLRCprime}). }
\end{figure*}

A closeup of the low-frequency region is given in Fig. \ref{fig9}, comparing DGSS with linear-response calculations using the xc kernel of
Eq. (\ref{fxcLRCprime}) with values of $\gamma$ as indicated. In addition to this, we also choose a finite damping term $\beta=0.01$ to avoid
a singularity if $\omega^2 = \gamma$. For $\alpha=5.0$ and 7.0 the two methods are in excellent agreement, apart from a small frequency shift
of the low-frequency signal. This small discrepancy is because the modified xc kernel (\ref{fxcLRCprime}) is strictly valid only in the linear regime, and the real-time
propagation unavoidably introduces some nonlinearities. These nonlinearities become more significant for larger $\alpha$ due to the larger oscillator
strength of the exciton; hence, a greater shift for larger $\alpha$.
For $\alpha=1$ and 3, there is a difference between the shapes of the features around $\omega=0.1$ comparing
the real-time and linear-response calculations. The likely reason for this is that the real-time calculation uses a sudden switch-on
at time $t=0$, which affects the phase of the subsequent dipole oscillations.

We observe in Fig. \ref{fig9} that the imaginary part of $\epsilon(\omega)$ dips below zero at the low-frequency resonances, which indicates
anti-absorptive behavior -- in other words, the system acts as a source. As we will explain below when we discuss the DGSS stabilization mechanism, this unexpected
behavior comes from the fact that the $\gamma$ parameter causes an effective counter force to suppress the runaway oscillations.

\subsection{TDLRC as a parametric oscillator}\label{Sec:4C}

As the above results demonstrate, the DGSS approach very effectively stabilizes the real-time propagation in the presence of
LRC-type excitonic binding. The main concern is now how this approach can be justified. To arrive at an answer, in this and the next subsection we
formulate two questions and then show how they are closely related to each other.

The first question is this: how is the excitonic resonance generated in a real-time propagation? From linear response theory we know that
excitons are a collective phenomenon in which individual particle-hole pairs join together to form a collective excitation. But how does it work in real time?

It is helpful to consider a classical analog: the parametric oscillator. A parametric oscillator is a dynamical system which obeys an equation of
motion of the generic type
\begin{equation}\label{parametric_oscillator}
\ddot u(t) + \omega^2(t) u(t) = 0 \:,
\end{equation}
where $\omega(t)$ can be viewed as a generalized ``spring constant'', which describes some type of external force or driver that is capable of shaking up the system.
A common form is $\omega^2(t) = a + b \cos t$, which leads to the Mathieu equation which has been extensively discussed in the literature (here, $a,b$ are real constants)
\cite{Kovacic2018}.
Parametric oscillators of this type are characterized by regions of stability, featuring steady periodic behavior; but there are also regions of
instability.

\begin{figure*}
  \includegraphics[width=\linewidth]{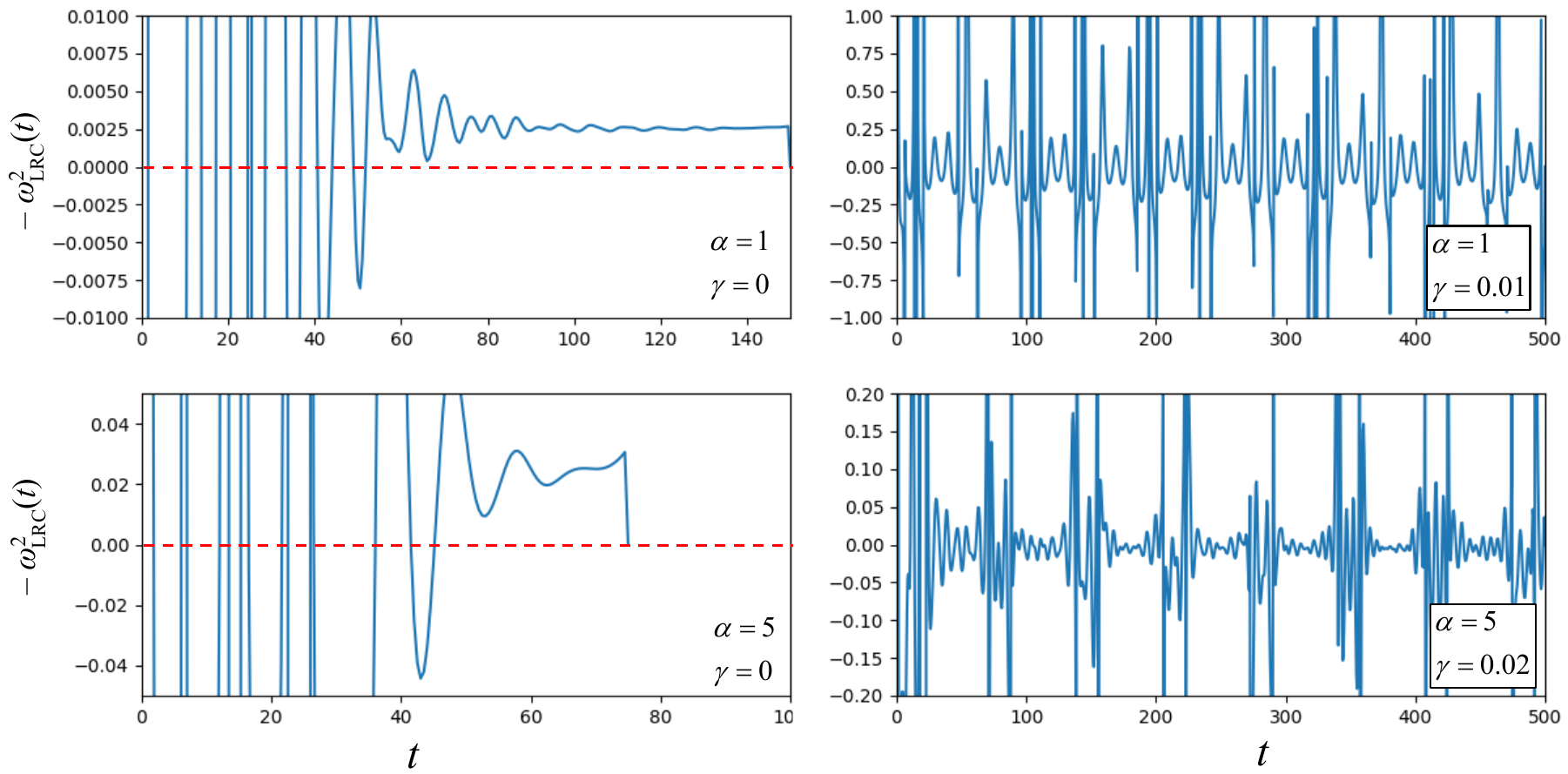}
  \caption{\label{fig10} Generalized spring constant $-\omega^2_{\rm LRC}(t)$ of TDLRC parametric oscillations, for $\alpha=1.0$ and 5.0. If $\gamma=0$,
  TDLRC is unstable, and the generalized spring constant has a finite offset (left panels). If $\gamma$ is chosen finite to stabilize the calculations,
  then $-\omega^2_{\rm LRC}(t)$ varies about zero (right panels).  }
\end{figure*}

The parallel between classical parametric oscillators and the TDLRC approach to generate excitons is quite clear. The Kohn-Sham
system is driven by the time-dependent xc vector potential $\bfA_{\rm xc,\bfzero}^{\rm LRC}$, which obeys the equations of motion
(\ref{LRC-vec-pot-diffeq-3D}) or (\ref{LRC-vec-pot-diffeq-2D}), depending on the dimension of the system.
The macroscopic current density $\bfj_\bfzero(t)$ on the right-hand side
is a complicated functional of $\bfA_{\rm xc,\bfzero}^{\rm LRC}$: the dependence is explicitly through Eq. (\ref{jmac}) and also implicit, since
the xc vector potential enters the TDKS equation, which in turn determines the single-particle wave functions. However, we can {\em define}
 a corresponding parametric oscillator as
\begin{equation}
\frac{\ddot \bfA_{\rm xc,\bfzero}^{\rm LRC}(t)}{\bfA_{\rm xc,\bfzero}^{\rm LRC}(t)} = -\omega^2_{\rm LRC}(t) \:.
\end{equation}
Since $\bfA_{\rm xc,\bfzero}^{\rm LRC}(t)$ is known, $\omega^2_{\rm LRC}(t)$ can be numerically constructed. This is illustrated in Fig. \ref{fig10},
which shows the generalized spring constant $-\omega^2_{\rm LRC}(t)$ for two cases, $\alpha=1.0$ and 5.0. The calculations are unstable if $\gamma=0$:
in that case, $-\omega^2_{\rm LRC}(t)$  has a finite offset (the zero is indicated by the red dashed lines), which causes $ \bfA_{\rm xc,\bfzero}^{\rm LRC}(t)$
to diverge exponentially. By contrast, if $\gamma$ is chosen to be finite ($\gamma=0.01$ for $\alpha=1.0$ and $\gamma=0.02$ for $\alpha=5.0$)
then $-\omega^2_{\rm LRC}(t)$  has a strong oscillatory behavior, but is zero on average. Thus, the parametric oscillations are stable.
However, we still need an explanation for $\gamma$.

\subsection{The role of the zero-force theorem}\label{Sec:4D}

We will now discuss the zero-force theorem of TDDFT \cite{Ullrich2012} and show that this important theorem is violated by TDLRC.
The second of the two questions is then: how is this related to the observed instabilities, and can this lead to a cure?
We will show that it is possible to define a counter vector potential to enforce the zero-force theorem, and how this provides a
natural explanation for the success of the DGSS approach.

\subsubsection{Variationally optimized function satisfying a constraint}

The idea to enforce exact conditions on approximate xc functionals by constrained minimization
was first introduced by Kurzweil, Baer and Head-Gordon \cite{Kurzweil2008,Kurzweil2009}. The approach was recently used in the context of the
zero-torque theorem in spin-DFT \cite{Pluhar2019}.
In the following, we will use a variational approach to find the optimal counter vector potential to enforce the zero-force theorem
(the meaning of ``optimal'' will be defined below). 
In this subsection we will consider a simple mathematical problem to motivate and explain our approach; based on this,
the counter vector potential will then be constructed in the following subsection.

Assume that a well-behaved positive function $n(x)$ is given, which is normalized to 1 over an interval:
\begin{equation}
n(x)>0\:, \qquad \qquad \int_{-L}^L n(x)dx= 1 \:.
\end{equation}
Now find a function $f(x)$ with the property
\begin{equation}
\int_{-L}^L n(x) f(x)dx= F \:.
\end{equation}
The immediate trivial solution is $f(x) = F$. But there are infinitely many other solutions, and we are particularly interested
in that solution whose norm is minimal. Let us formulate this as a minimization of a functional $J$,
\begin{equation}
J = \int_{-L}^L f^2(x)dx \equiv ||f^2|| \:,
\end{equation}
under the constraint
\begin{equation} \label{C4}
g = \int_{-L}^L n(x) f(x)dx -F =0\:.
\end{equation}
The Euler-Lagrange equation for this is
\begin{equation}
\frac{\delta J}{\delta f(x)} + \lambda \frac{\delta g}{\delta f(x)} = 0 \:,
\end{equation}
which leads to
\begin{equation}
f(x) = -\frac{\lambda}{2} n(x) \:.
\end{equation}
We determine the Lagrange multiplier $\lambda$ by substituting into Eq. (\ref{C4}),
which gives the final result
\begin{equation}
f(x) = \frac{F n(x)}{||n^2||} \:.
\end{equation}
As an example, consider $n(x) = e^{-x^2}/\sqrt{\pi}$, for $L\to\infty$. We find
\begin{equation}
f(x) = \frac{Fe^{-x^2}/\sqrt{\pi}}{\int (e^{-2x'^2}/\pi) dx'} = \sqrt{2} F e^{-x^2} \:,
\end{equation}
which gives $||f^2|| = \sqrt{2\pi} F^2$.
By comparison, the trivial solution $f(x) = F$ gives $||f^2|| = 2LF^2 \to \infty$.

\subsubsection{Zero-force theorem and LRC counter-force}

The zero-force theorem for the scalar xc potential reads
\begin{equation}
\int d\bfr \: n(\bfr,t) \nabla v_{\rm xc}(\bfr,t) = 0 \:.
\end{equation}
Now imagine that we have an approximate xc potential that does not satisfy the zero-force theorem, but yields a finite total force:
\begin{equation}
\int  d\bfr \:n(\bfr,t) \nabla v_{\rm xc}^{\rm approx}(\bfr,t) =  \bfF(t)\:.
\end{equation}
We want a counter-force $\bff(\bfr,t)$ that restores the zero-force theorem, and which has a vector norm that is as small as possible:
\begin{equation}
\int  d\bfr \:n(\bfr,t) \bff(\bfr,t) =  -\bfF(t) \:,
\end{equation}
where
\begin{equation}
||\bff^2(t)|| = \int d\bfr\: |\bff(\bfr,t)|^2 = min \:.
\end{equation}
In complete analogy with the warm-up exercise of the preceding subsection, we find
\begin{equation}
\bff(\bfr,t) = -\frac{\bfF(t) n(\bfr,t)}{\int d\bfr \: n^2(\bfr,t)} \:.
\end{equation}
Can this counter-force be written as the gradient of a scalar counter-potential, $\bff(\bfr,t) = -\nabla v_{\rm xc}^{\rm counter}(\bfr,t)$?
Since $\bff(\bfr,t)$ might not be curl free, the general answer is no.
However, we can gauge transform it into a counter-vector-potential:
\begin{equation}
\frac{\partial}{\partial t}\bfA_{\rm xc}^{\rm counter}(\bfr,t) = -\frac{\bfF(t) n(\bfr,t)}{|| n^2(t)||} \:.
\end{equation}
For a spatially periodic system, this becomes
\begin{equation}
\bfA^{\rm counter}_{\rm xc,\bfG}(t) = -\int_0^t dt' \frac{\bfF(t') n_\bfG(t')}{|| n^2(t')||} \:.
\end{equation}
Now let us connect this to our case, where the zero-force theorem is violated by the macroscopic LRC vector potential. Thus, for the 3D case (and similar in 2D),
\begin{equation}
\bfF(t) = \alpha N \int_0^t dt' \: \bfj_\bfzero(t')\:.
\end{equation}
We end up with
\begin{equation}\label{AG}
\bfA^{\rm counter}_{\rm xc,\bfG}(t) = -\alpha N \int_0^t dt' \frac{ n_\bfG(t')}{|| n^2(t')||}\int_0^{t'} dt'' \: \bfj_\bfzero(t'') \:.
\end{equation}

In particular, the macroscopic counter vector potential is, using $n_\bfzero = N/{\cal V}$,
\begin{equation}\label{C18}
\bfA^{\rm counter}_{\rm xc,\bfzero}(t) = -\frac{\alpha N^2}{\cal V} \int_0^t dt' \frac{1}{|| n^2(t')||}\int_0^{t'} dt'' \: \bfj_\bfzero(t'') \:.
\end{equation}
Comparing this to the macroscopic LRC vector potential $\bfA^{\rm LRC}_{\rm xc,\bfzero}(t)$, Eq. (\ref{A3D}), we see that
the macroscopic counter term will be smaller than the macroscopic LRC term
if $|| n^2(t')|| > \frac{N^2}{\cal V}$, which is usually the case.

\begin{figure}
  \includegraphics[width=\linewidth]{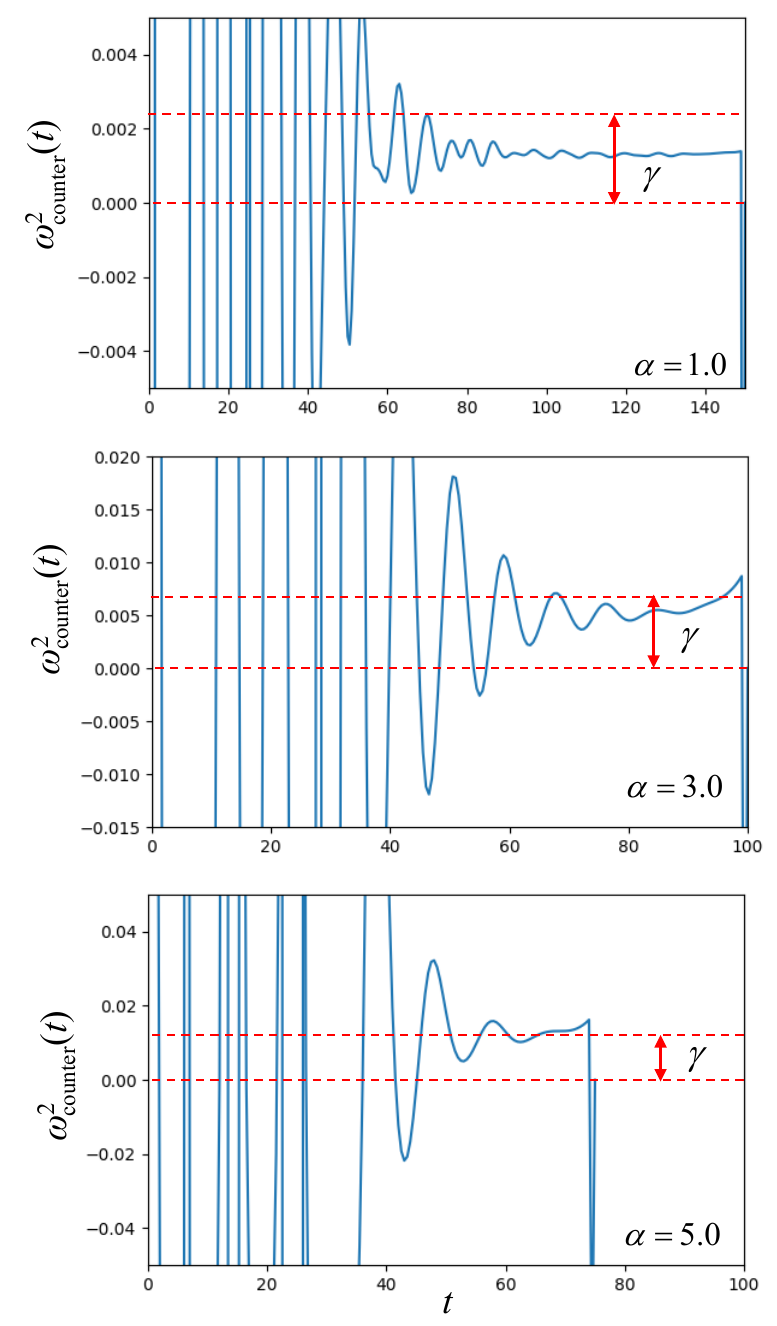}
  \caption{\label{fig11} Generalized spring constant $\omega^2_{\rm counter}(t)$ associated with the TDLRC counter vector potential, see Eq. (\ref{omega_counter}).
  The offset is identified with the $\gamma$-parameter that is required to stabilize the calculation. For $\alpha=1.0$ we find $\gamma = 0.0024$, for
  $\alpha=3.0$ we find $\gamma = 0.0068$,  and for $\alpha=5.0$ we find $\gamma = 0.012$.}
\end{figure}

\subsubsection{Stabilization of TDLRC via the averaged counter force}

Let us now use the results derived above, in particular the macroscopic counter term $\bfA_{\rm xc,\bfzero}^{\rm counter}(t)$ for TDLRC given by Eq. (\ref{C18}),
and explain how the violation of the zero-force theorem is related to the observed instabilities, and how the counter force can provide a remedy.

We use the following insight: to stabilize the TDLRC calculations it is necessary that the macroscopic net force on the system is zero {\em on average}.
Let us define the generalized parametric spring constant associated with the counter vector potential:
\begin{equation}\label{omega_counter}
\frac{\ddot \bfA_{\rm xc,\bfzero}^{\rm counter}(t)}{\bfA_{\rm xc,\bfzero}^{\rm LRC}(t)} = \omega^2_{\rm counter}(t) \:.
\end{equation}
This quantity is plotted in Fig. \ref{fig11}; it is the same as in the left panels of Fig. \ref{fig10}, but slightly downscaled.
As indicated, we identify the offset with the smallest $\gamma$-parameter that is necessary to stabilize TDLRC. For $\alpha=1.0$ we find $\gamma = 0.0024$,
for $\alpha=3.0$ we find $\gamma = 0.0068$,
and for $\alpha=5.0$ we find $\gamma = 0.012$, in agreement with the results shown in Fig. \ref{fig7}.
In other words, we identify
\begin{equation}
\gamma_{\rm thr} = \langle \omega^2_{\rm counter}(t) \rangle.
\end{equation}
This argument explains why there is a rather sharp transition between unstable and stable as the $\gamma$-parameter is
increased from zero: at the threshold and beyond, the zero-force theorem is satisfied on average.

Unfortunately, this explanation does not have predictive capabilities: $\gamma_{\rm thr}$ depends
on the system parameters and on the details of the numerical implementation, but a mathematically more rigorous stability analysis
appears difficult. In practice, however, this is not a critical problem,
as already noted by  DGSS \cite{Dewhurst2024}, who found the $\gamma$ parameter to be largely material independent.
This is in line with our findings: once $\gamma$ has passed the threshold value required for stabilization, there is a
rather broad range of values of $\gamma$ that leads to stable dipole oscillations and that will produce essentially the
same optical spectra.

One might wonder why we do not simply take the time-dependent $\bfA_{\rm xc,\bfzero}^{\rm counter}(t)$ and put it directly into the TDKS equation (\ref{TDKSk}).
In the limit of small perturbations, $||n^2(t)||$ remains nearly constant and can then be pulled out of the integral over $dt'$ in
Eq. (\ref{C18}). This means that
\begin{equation}\label{C18approx}
\bfA^{\rm counter}_{\rm xc,\bfzero}(t) \approx - \alpha_{\rm counter} \int_0^t dt' \int_0^{t'} dt'' \: \bfj_\bfzero(t'') \:,
\end{equation}
which has the same form as $\bfA_{\rm xc,\bfzero}^{\rm LRC}(t)$ from Eq. (\ref{A3D}) apart from the different prefactor $\alpha_{\rm counter} = \alpha N^2/({\cal V} ||n^2||)$ and the overall minus sign. Directly combining the LRC and counter vector potentials would thus effectively make $\alpha$ smaller,
which would of course have a stabilizing effect by simply decreasing the excitonic binding energy, but this is not what we want.
We also briefly mention that we tested the microscopic (i.e., finite-$\bfG$) contributions to the counter vector potential, $\bfA_{\rm xc,\bfG}^{\rm counter}(t)$,
and found their effect to be rather small.

\subsection{Additional properties of DGSS}
\subsubsection{Connection between $\gamma_{\rm thr}$ and exciton characteristics}

\begin{figure}
  \includegraphics[width=\linewidth]{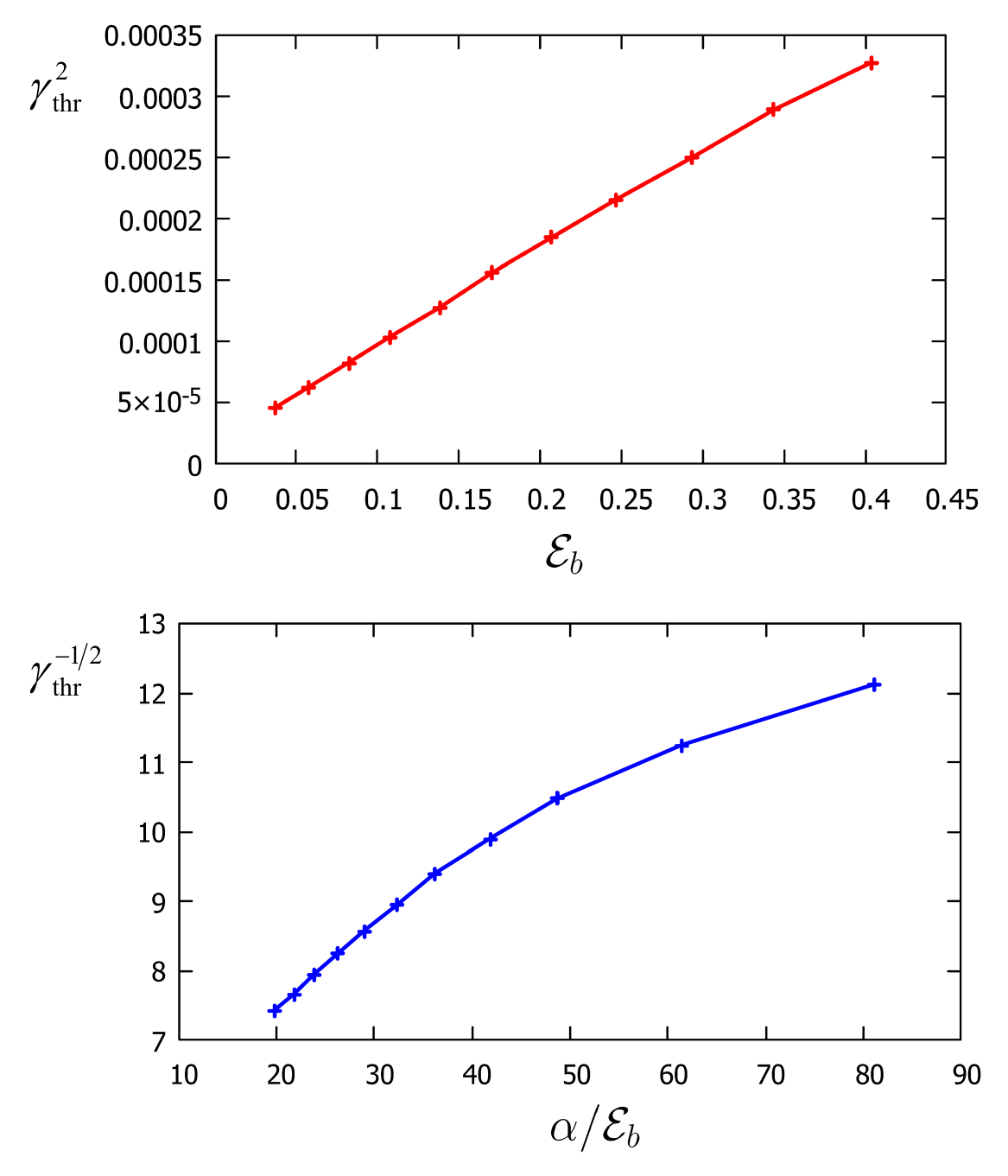}
  \caption{\label{fig12} Top: square of the threshold value of the $\gamma$-parameter, $\gamma_{\rm thr}^2$, versus exciton binding energy ${\cal E}_b$.
  Bottom: $1/\sqrt{\gamma_{\rm thr}}$ versus $\alpha/{\cal E}_b$, which is estimated to be proportional to the exciton Bohr radius within the 2D Wannier model.}
\end{figure}

According to Ref. \cite{Dewhurst2024}, in the DGSS approach there is a natural length scale $1/\sqrt{\gamma}$ which is linearly related to the exciton
Bohr radius; DGSS showed that this is consistent with numerical data for several materials of interest, with the exception of the outlier GaAs.

We find similar trends in our 2D model system, which may offer some ways to predict the values of $\gamma$ required for stabilization.
The top panel of Fig. \ref{fig12} shows that there is an almost perfect linear relation between the exciton binding energy ${\cal E}_b$ and
the square of the threshold value of the $\gamma$-parameter, i.e., ${\cal E}_b \sim \gamma_{\rm thr}^2$. The exciton binding energy can be obtained from Fig. \ref{fig4},
taking the difference between the band gap and the exciton peak positions.

Based on the 2D Wannier model \cite{Haug} for the exciton radius $a_0$, one arrives at the simple estimate $a_0 \sim \alpha/{\cal E}_b$.
This assumes that the LRC parameter behaves as $\alpha\sim \varepsilon^{-1}$ \cite{Botti2004}, where here $\varepsilon$ is the background dielectric constant
that enters in the 2D Wannier model \cite{Haug}.
The bottom panel of Fig. \ref{fig12} plots $1/\sqrt{\gamma_{\rm thr}}$ versus $\alpha/{\cal E}_b$ and finds indeed a linear relationship for small exciton radii,
as predicted by DGSS. We also find strong deviations from this linear relationship for large exciton radii (weakly bound excitons),
which is consistent with the outlier behavior of GaAs \cite{Dewhurst2024}.

\subsubsection{Nonlinear regime}

Lastly, we address the question whether the introduction of the $\gamma$-parameter in the DGSS approach ensures stability
even beyond the linear regime. DGSS found that this was indeed the case, which allowed them to study bleaching of the excitonic response in bulk silicon
upon strong laser excitation \cite{Dewhurst2024}.

\begin{figure*}
  \includegraphics[width=\linewidth]{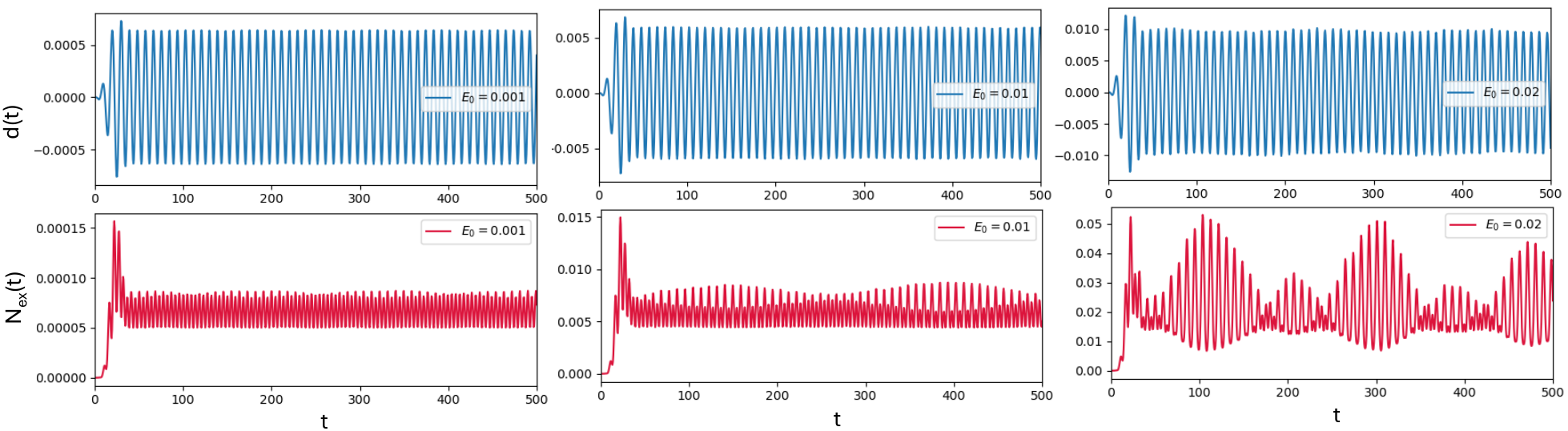}
  \caption{\label{fig13} Time-dependent dipole moment $d(t)$ and number of excited electrons per unit cell $N_{\rm ex}(t)$ for $\alpha=4$ and $\gamma = 0.009$.
  The system is excited by 3-cycle pulses with $\omega=0.5$ and peak electric field strength $E_0 = 0.001$ (left), $E_0 = 0.01$ (middle), and $E_0 = 0.02$ (right).}
\end{figure*}

We simulate laser excitation of our 2D model solid with a short, 3-cycle pulse  with $\sin^2$ envelope and peak electric field strength $E_0$, linearly polarized along the
$x-y$ diagonal. We consider the case $\alpha = 4$, which gives a strong excitonic peak with $E_b = 0.082$, see Fig. \ref{fig4}. The laser pulse has frequency $\omega = 0.5$,
i.e., below-gap and off-resonance with the exciton. Results are shown in Fig. \ref{fig13} for three peak field strengths: $E_0 = 0.001$, 0.01 and 0.02 a.u., which
corresponds to the peak intensities $I=3.51\times 10^{10}$, $3.51\times 10^{12}$, and $1.40\times 10^{13}\: \rm W/cm^2$, respectively.

In all three cases shown in Fig. \ref{fig13} the time-dependent dipole moment $d(t)$ (upper panels) is stable using $\gamma=0.009$. The lower panels show the number of excited electrons
per unit cell, $N_{\rm ex}(t)$, which is obtained by projecting the TDKS orbitals onto the ground-state band structure. For the lowest intensity,
less than $0.0001$ electrons per unit cell are promoted to the conduction band, which is a weak excitation. For  $E_0 = 0.01$, we find $N_{\rm ex} \approx 0.006$
on average, and for $E_0 = 0.02$ we find $N_{\rm ex} \approx 0.02$ on average, with rather strong fluctuations. The latter case means that on the order of 1\% of electrons per unit
cell are excited into the conduction band, which is significant and puts us already into the strongly nonlinear regime.

For $E_0 \ge 0.024$ ($I\ge 2\times 10^{13}\: \rm W/cm^2$), the calculations become unstable within DGSS, for any choice of $\gamma$. This corresponds to an extremely
nonlinear regime in which $N_{\rm ex}>0.1$. Thus, for the case considered here, we conclude that the DGSS approach remains stable over a wide range of intensities
up until the extremely nonlinear regime. Further studies will be needed to explore the intensity dependence of DGSS more systematically, which is beyond the scope
of this paper.

\section{Conclusions}\label{Sec:5}

In this paper we have examined the performance of TDDFT for calculating optical absorption spectra in insulators and
semiconductors, using the LRC approach. Earlier work \cite{Sun2021} had shown that TDLRC can successfully simulate
weakly bound excitons, but fails for strong electron-hole interactions. Recently, DGSS \cite{Dewhurst2024}
proposed a simple method, termed the Kohn-Sham-Proca scheme, to stabilize the TDLRC calculations, which they used to successfully calculate optical spectra
with strong excitonic effects.

We have used a 2D model solid to investigate the reasons for the numerical instabilities of TDLRC and to explain how and why
the method of DGSS works. The answer reveals the key role played by the zero-force theorem, and gives
important new insight in the way excitons are produced using the TDLRC approach. To generate excitonic resonances,
the xc vector potential must drive the system with the corresponding frequencies. In TDLRC, this can be viewed as
a parametric oscillator, with a time-dependent spring constant that ensures the correct driving frequency for the vector potential.
However, parametric oscillations are prone to instabilities; the $\gamma$-parameter of the DGSS scheme ensures that
the parametric oscillator remains in a stable region. The required choice of $\gamma$ is such that the zero-force theorem
is satisfied on average. Importantly, trying to enforce the zero-force theorem at each time $t$ would spoil the TDLRC method;
the key point is that it holds on average. We also found a simple linear relationship between the threshold value of $\gamma$
and the exciton binding energy, which may offer at least some degree of predictability.

These findings have several implications. On the practical side, TDLRC is a numerically cheap method that can describe
excitonic effects in real time, and should therefore be very useful to describe a variety of nonequilibrium phenomena
in semiconductors and insulators. Strictly speaking, the TDLRC approach defined here is only valid in the limit of
weak perturbations, where we have shown that it agrees with linear-response LRC. However, as long as the system is
not driven too strongly, TDLRC (and DGSS) should remain reasonably accurate even beyond the linear response limit; as we have demonstrated,
DGSS remains numerically stable far into the nonlinear regime. Applications
to ultrafast, strongly driven processes or transient absorption phenomena are therefore within reach, as shown in Ref. \cite{Dewhurst2024}.

A more fundamental question concerns the role of the zero-force theorem and other exact constraints \cite{zerotorquenote} in TDDFT.
We have seen here that physically meaningful and practically useful results
can follow from an xc functional that manifestly violates the zero-force theorem, as long as these violations vanish in the
time average. It would be interesting to explore the consequences of this for describing other types of collective excitations such
as plasmons or magnons, or for ``fixing'' other types of approximate xc functionals in TDDFT \cite{Mundt2007}.
Ultimately, though, the task will be to obtain a TDDFT approach that is capable of capturing the
long-range electron-hole interaction that gives rise to excitonic physics, without running in conflict with the zero-force theorem
and valid in the linear as well as nonlinear regime.
The methods discussed here could be a good starting point to discover such approximations.

\acknowledgments

This work was supported by NSF Grant No. DMR-2149082. We acknowledge helpful discussions with Sangeeta Sharma, Kay Dew\-hurst, Nicolas Tancogne-Dejean,
and Yifan Yao.

\appendix

\section{Dielectric function from linear-response TDDFT with the LRC kernel}\label{App:A}

The proper density-density response function $\tilde \chi$ is defined as follows:
\begin{eqnarray} \label{M.proper}
\tchi(\bfr,\bfr',\omega) &=&  \chi_0(\bfr,\bfr',\omega)
+ \int d\bfr_1 \int d\bfr_2 \: \chi_0(\bfr,\bfr_1,\omega) \nonumber\\
&& \times
f_{\rm xc}(\bfr_1,\bfr_2)\tchi(\bfr_2,\bfr',\omega) \:,
\end{eqnarray}
where $\chi_0$ is the response function of the noninteracting Kohn-Sham system and $f_{\rm xc}$ is the xc kernel, assumed here to be frequency-independent.
With this, we can express the dielectric function as
\begin{equation} \label{M.e}
\epsilon(\bfr,\bfr',\omega) = \delta(\bfr - \bfr') - \int d\bfr'' \: \frac{\tchi(\bfr'',\bfr',\omega)}{|\bfr - \bfr''|} \:.
\end{equation}
Replacing $\tchi$ with $\chi_{0}$ defines the random-phase approximation (RPA) dielectric function; here, however, we wish
to go beyond the RPA.

For lattice-periodic systems, Eq. (\ref{M.e}) turns into
\begin{equation}
\epsilon_{\bfG \bfG'}(\bfk,\omega) = \delta_{\bfG \bfG'} - w_{\bfk-\bfG} \tchi_{\bfG \bfG'}(\bfk,\omega) \:,
\label{e_rec}
\end{equation}
where $w_{\bfk-\bfG}$ is defined in Eq. (\ref{w}).
The reciprocal-space form of the  Dyson-like equation (\ref{M.proper}) is
\begin{eqnarray}\label{Dyson}
\tchi_{\bfG \bfG'}(\bfk,\omega) &=& \chi_{0,\bfG \bfG'}(\bfk,\omega)
+ \sum_{\bfG_1 \bfG_2} \chi_{0,\bfG \bfG_1}(\bfk,\omega) \nonumber\\
&& \times
f^{\rm xc}_{\bfG_1\bfG_2}(\bfk,\omega) \tchi_{\bfG_2 \bfG'}(\bfk,\omega)\:,
\end{eqnarray}
where the LRC xc kernel [see Eq. (\ref{fxcLRC})] is given by \cite{Byun2017b}
\begin{equation}
f^{\rm xc}_{\bfG_1\bfG_2}(\bfk,\omega) = -\frac{\alpha}{4\pi} w_{\bfk-\bfG_1} \delta_{\bfG_1\bfG_2} \:.
\end{equation}
We are interested in the macroscopic dielectric function,
\begin{equation}\label{emac}
\epsilon_{\rm mac}(\omega) = 1 - \lim_{\bfk\to 0} w_\bfk \tchi_{00}(\bfk,\omega) \:.
\end{equation}
The calculation of $\tchi_{00}(\bfk,\omega)$  can be achieved by setting $\bfG'=\bfzero$ in Eq. (\ref{Dyson}), which then leads to
\begin{eqnarray}\label{Dyson2}
\lefteqn{\tchi_{\bfG \bfzero}(\bfk,\omega) = \chi_{0,\bfG \bfzero}(\bfk,\omega)} \nonumber\\
&&{}-
\frac{\alpha}{4\pi} \sum_{\bfG_1 } \chi_{0,\bfG \bfG_1}(\bfk,\omega)
w_{\bfk-\bfG_1} \tchi_{\bfG_1 \bfzero}(\bfk,\omega) \:.
\end{eqnarray}
This can be cast into a system of linear equations, which can be solved with moderate numerical effort.
To be specific, we obtain
\begin{eqnarray}\label{Dyson5}
\lefteqn{ \hspace{-2.5cm} \sum_{\bfG_1 } \left[
\delta_{\bfG\bfG_1} + \frac{\alpha}{4\pi}  w_{\bfk-\bfG_1} \chi_{0,\bfG \bfG_1}(\bfk,\omega) \right]\tchi_{\bfG_1 \bfzero}(\bfk,\omega)}\nonumber\\
&&{}= \chi_{0,\bfG \bfzero}(\bfk,\omega) \:.
\end{eqnarray}

The noninteracting response function for periodic systems is given by
\begin{eqnarray}\label{chi}
\chi_{0,\bfG \bfG'}(\bfk,\omega) &=&
\frac{2}{\cal V} \sum_{\bfk'}\sum_{j,l=1}^\infty
\frac{f_{j\bfk'} - f_{l \bfk + \bfk'}}{\omega + \varepsilon_{j\bfk'} - \varepsilon_{l\bfk + \bfk'} + i\eta}\nonumber\\
&&{}\times
\langle j\bfk' | e^{-i(\bfk + \bfG)\cdot\bfr} | l\bfk + \bfk' \rangle \nonumber\\
&&{}\times
\langle l\bfk + \bfk' |  e^{i(\bfk + \bfG')\cdot\bfr'} | j\bfk' \rangle \:.
\end{eqnarray}
In the following, we need the $\bfk\to 0$ limit of this, which is given by
\begin{eqnarray}
\lefteqn{\chi_{0,\bfG \bfG'}(\bfk\to 0,\omega)}\nonumber\\
&=&
\frac{2}{\cal V} \sum_{\bfk'}\sum_j^{\rm occ} \sum_{l}^{\rm empty}
\frac{\langle j\bfk' | e^{-i(\bfk + \bfG)\cdot\bfr} | l\bfk' \rangle
\langle l\bfk' |  e^{i(\bfk + \bfG')\cdot\bfr} | j\bfk' \rangle }
{\varepsilon_{j\bfk'} - \varepsilon_{l\bfk'} + \omega + i\eta}
\nonumber\\
&+&
\frac{2}{\cal V} \sum_{\bfk'}\sum_j^{\rm occ} \sum_l^{\rm empty}
\frac{\langle j\bfk' |  e^{-i(\bfk + \bfG')\cdot\bfr} | l\bfk' \rangle
\langle l\bfk' | e^{i(\bfk + \bfG)\cdot\bfr} | j\bfk' \rangle}
{ \varepsilon_{j\bfk'} - \varepsilon_{l\bfk'}  - \omega - i\eta} \:.\nonumber\\
&&
\end{eqnarray}
For finite reciprocal lattice vectors we can set directly $\bfk=0$ and then obtain matrix elements of the form
\begin{eqnarray}
\lefteqn{\hspace{-1cm} \langle j\bfk' | e^{-i\bfG\cdot\bfr} | l\bfk' \rangle
=
\int d\bfr\:  u^*_{j\bfk'}(\bfr)  e^{-i\bfG\cdot\bfr} u_{l\bfk'}(\bfr)} \nonumber\\
&=&
\sum_{\bfG_1 \bfG_2} C^*_{j \bfk'-\bfG_1}C_{l \bfk'-\bfG_2} \delta_{\bfG,\bfG_1-\bfG_2} \:.
\end{eqnarray}
For $\bfG=0$ or $\bfG'=0$, we have to deal with matrix elements of the form
$\langle j\bfk' | \bfk\cdot\bfr | l\bfk' \rangle$, which reduces to the evaluation of the dipole matrix elements ${\bm \mu}_{jl\bfk'}$, see Eqs. (\ref{mu1}) and (\ref{mu2}).

A considerable simplification can be achieved by making
the head-only approximation of the LRC kernel \cite{Byun2017b}:
\begin{equation}
f^{\rm xc}_{\bfG_1\bfG_2}(\bfk,\omega) \approx -\frac{\alpha}{4\pi} v_{\bfG_1}(\bfk) \delta_{\bfG_1\bfG_2}\delta_{\bfG_1 \bfzero} \:,
\end{equation}
so that Eq. (\ref{Dyson5}) becomes
\begin{equation}\label{Dyson3}
\tchi_{\bfzero \bfzero}(\bfk,\omega) = \chi_{0,\bfzero \bfzero}(\bfk,\omega)
-\frac{\alpha}{4\pi}  \chi_{0,\bfzero \bfzero}(\bfk,\omega) w_\bfk \tchi_{\bfzero \bfzero}(\bfk,\omega)
\end{equation}
or
\begin{equation} \label{Dyson4}
\tchi_{\bfzero \bfzero}(\bfk,\omega) = \frac{\chi_{0,\bfzero \bfzero}(\bfk,\omega)}{ 1 + \frac{\alpha}{4\pi} w_\bfk \chi_{0,\bfzero \bfzero}(\bfk,\omega) } \:.
\end{equation}
Again, in 2D the limit $\bfk\to 0$ is understood as a small but finite value of $\bfk$, since $\chi_{0,\bfzero \bfzero}(\bfk,\omega) $ goes to zero as
$k^2$, but $w_\bfk$ only diverges as $k^{-1}$. In 3D, on the other hand, there is an exact cancellation of $k^2$.

\section{Dielectric function from real-time TDDFT}\label{App:B}

We follow Sander and Kresse \cite{Sander2017} and calculate the macroscopic dielectric function via the dipole polarizability. Starting point is the following expression for the time-dependent dipole moment:
\begin{equation}
    {\bf d}(t) = \int_{\rm cell} \bfr\, n(\bfr,t) d\bfr \:.
\end{equation}
However, the dipole operator is ill defined in periodic systems. To find an alternative expression for ${\bf d}(t)$, we expand the periodic part of the time-dependent Kohn-Sham Bloch functions as
\begin{equation}
    u_{l\bfk}(\bfr,t) = \sum_m \xi_{lm\bfk}(t) u_{m\bfk}^{(0)}(\bfr) \:,
\end{equation}
where the $u_{m\bfk}^{(0)}(\bfr)$ come from the static Kohn-Sham equation (\ref{KS}), and the expansion coefficients are given by
\begin{equation}
    \xi_{lm\bfk}(t) = {\cal V} \sum_\bfG C_{m,\bfk-\bfG}^{(0)} C_{l,\bfk-\bfG}(t) \:.
\end{equation}
The time-dependent dipole moment can then be written as
\begin{equation}\label{dmu}
    {\bf d}(t) = 2\sum_l^{N/2} \sum_\bfk \sum_{m\ne m'} \xi^*_{lm\bfk}(t) \xi_{lm' \bfk}(t) {\bm \mu}_{mm'\bfk} \:.
\end{equation}
The dipole matrix elements,
\begin{equation}\label{mu1}
    {\bm \mu}_{mm'\bfk} = \int_{\rm cell} u^{(0)*}_{m\bfk} \bfr u^{(0)}_{m'\bfk} d\bfr\:,
\end{equation}
are still ill defined at this stage. However, using the commutator between the ground-state Hamiltonian and the dipole operator, they can be recast into \cite{Sander2017}
\begin{equation}\label{mu2}
    {\bm \mu}_{mm'\bfk} = \frac{{\cal V} \sum_\bfG \bfG C^{(0)*}_{m,\bfk-\bfG} C^{(0)}_{m',\bfk-\bfG}}{\varepsilon_{m\bfk} - \varepsilon_{m'\bfk}} \: .
\end{equation}
This expression is now well behaved and can be used in Eq. (\ref{dmu}) to calculate ${\bf d}(t)$ at each time step.

The dynamical polarizability $\tilde\alpha(\omega)$ then follows from the Fourier transform
\begin{equation}
    \tilde\alpha(\omega) = \frac{1}{E_0} \int_{t_0}^{T} dt' d(t') e^{-i\omega t - \eta t} \:,
\end{equation}
where we assume for simplicity that the dipole moment ${\bf d}(t)$ has been evaluated along the same direction as the vector potential kick, Eq. (\ref{kick}). Here, $\eta$ is a small imaginary part of the frequency which introduces a numerical line broadening in the spectrum. The 2D macroscopic dielectric function then follows from this as
\begin{equation}\label{B8}
    \epsilon_{\rm mac}(\omega) = 1 - \lim _{k\to 0} 2\pi k \tilde\alpha(\omega) \:,
\end{equation}
to be evaluated at a small but finite wavevector.

We mention that we also implemented an alternative and in principle equivalent way to calculate $\epsilon_{\rm mac}$, namely via the macroscopic current density and the optical conductivity \cite{Yabana2012}. However, we found that for our system it is preferable to use the dipole moment as described here.

\bibliography{TDLRC_refs}

\begin{thebibliography}{58}
\bibitem [{cur()}]{currentnote}%
  \BibitemOpen
  \href@noop {} {}\bibinfo {note} {{To be precise, $\bfj(\bfr,t)$ is here the
  Kohn-Sham current density. Formally, the Kohn-Sham current density reproduces
  the exact physical current density if the exact xc scalar and vector
  potentials, $v_{\rm xc}(\bfr,t)$ and $\bfA_{\rm xc}(\bfr,t)$, are used in Eq.
  (\ref{TDKS}). In TDDFT without $\bfA_{\rm xc}(\bfr,t)$, only the longitudinal
  part of the Kohn-Sham current density can be in principle
  exact.}}\BibitemShut {Stop}%
\bibitem [{zer()}]{zerotorquenote}%
  \BibitemOpen
  \href@noop {} {}\bibinfo {note} {{Another exact constraint is the zero-torque
  theorem, related to conservation of angular momentum. Similar to the
  zero-force theorem, the long-range LRC vector potential will violate the
  zero-torque theorem, which might cause instability issues in chiral systems
  or under excitation with circularly polarized light. Whether this is
  mitigated by the DGSS approach remains to be seen.}}\BibitemShut {Stop}%
\end{thebibliography}
\end{document}